\documentclass[%
 reprint,
superscriptaddress,
nofootinbib,
 amsmath,amssymb,
 aps,
]{revtex4-2}

\usepackage{natbib} 
\usepackage{placeins}
\usepackage{times} 
\usepackage{amssymb,amsmath}
\usepackage{csquotes}
\usepackage{prelim2e}\usepackage[none,bottom]{draftcopy}
\draftcopyName{preprint / }{1.2} 
\ifx\pdfoutput\undefined
\usepackage[usenames,dvips]{color}
\usepackage{graphicx}     
\else
\usepackage[usenames,dvipsnames]{color}
\usepackage{epstopdf}
\usepackage[pdftex]{graphicx}
\usepackage[pdftex]{hyperref}
\hypersetup{a4paper,
    pdftitle={Manuscript on XXX} 
    pdfauthor={et al. and Uwe Thiele},  
pdfproducer={lateX},
pdfview=FitV,       
pdfstartview=FitB,
linkcolor=blue,     
citecolor=blue,     
urlcolor=red,      
breaklinks=true,    
colorlinks=true,
citebordercolor=0 0 0,  
filebordercolor=0 0 0,
linkbordercolor=0 0 0,
menubordercolor=0 0 0, 
urlbordercolor=0 0 0,
pdfhighlight=/I,
pdfborder=0 0 0,   
bookmarksopen=true,
bookmarksnumbered=true
}
\usepackage{pbox}
\fi
\ifx\pdfoutput\undefined
\DeclareGraphicsExtensions{.eps}
\else
\DeclareGraphicsExtensions{.jpg, .pdf, .tif, .png}
\fi
\usepackage{rotating}
\graphicspath{{.},{./Figures/}} 
\usepackage[usenames,dvipsnames]{color}
\usepackage[normalem]{ulem}
\renewcommand{\Re}[0]{\mathrm{Re\,}}
\renewcommand{\Im}[0]{\mathrm{Im\,}}
\newcommand{\vecg}[1]{\boldsymbol{#1}}
\newcommand{\tens}[1]{\mathbf{\underline{#1}}}

\newcommand{\Exp}[1]{\, \, \text{exp}\left(#1\right)}
\begin{document}
%
\title{
Nonreciprocal Cahn-Hilliard model emerges as a universal amplitude equation
}
\author{Tobias Frohoff-Hülsmann}
\email{t\_froh01@uni-muenster.de}
\thanks{ORCID ID: 0000-0002-5589-9397 }
\affiliation{Institute of Theoretical Physics, University of M\"unster, Wilhelm-Klemm-Str.\ 9, 48149 M\"unster, Germany}
\author{Uwe Thiele}
\email{u.thiele@uni-muenster.de}
\homepage{http://www.uwethiele.de}
\thanks{ORCID ID: 0000-0001-7989-9271}
\affiliation{Institute of Theoretical Physics, University of M\"unster, Wilhelm-Klemm-Str.\ 9, 48149 M\"unster, Germany}
\affiliation{Center for Nonlinear Science (CeNoS), University of M\"unster, Corrensstr.\ 2, 48149 M\"unster, Germany}
\begin{abstract}
  Oscillatory behavior is ubiquitous in out-of-equilibrium systems
  showing spatio-temporal pattern formation. Starting from a
  linear large-scale oscillatory instability -- a conserved-Hopf
  instability -- that naturally occurs in many active systems
  with two conservation laws, we derive a corresponding
  amplitude equation. It belongs to a hierarchy
  of such universal equations for the eight types of instabilities in homogeneous isotropic systems resulting from the combination of three features: large-scale vs.\ small-scale instability, stationary vs.\ oscillatory instability, and instability without and with conservation law(s).  The derived universal equation generalizes a phenomenological model of considerable recent interest, namely, the nonreciprocal Cahn-Hilliard model, and may be of a similar relevance for the classification of pattern forming systems as the complex Ginzburg-Landau equation.
\end{abstract}

\maketitle

The concept of active systems emerged as a paradigm in the description of a wide variety of biochemophysical nonequilibrium phenomena on multiple scales ranging from the collective behavior of molecules within biological cells to the dynamics of tissues or human crowds \cite{FKLW2019rmp}. In a narrow interpretation, active matter always involves chemo-mechanical coupling and shows some kind of self-sustained (collective) motion of the microscopic agents \cite{MJRL2013rmp,WaQX2021sm,SWWG2020nrp,BFMR2022prx}. In a wider sense, active systems encompass open systems that are kept out of equilibrium by a troughflow of material or energy \cite{Mikhailov1999}, and therefore may develop self-organized spatio-temporal patterns. This then includes the large spectrum of systems described by reaction-diffusion models \cite{PuBA2010ap,Liehr2013,KoDE2021ptrsapes} and systems characterized by the interplay of phase separation and chemical reactions \cite{BeBH2018rpp}.

In this context, predator-prey-type nonreciprocal interactions between constituents of active matter have recently become a particular focus as the implied breaking of Newton's third law results in a rich spectrum of nascent self-excited dynamic behavior \cite{ChKo2014jrsi,IBHD2015prx,KrIY2018sm,NaGo2020prl,FHLV2021n}. Beside various (stochastic) agent-based models of Langevin-type also continuous deterministic field theories have been proposed \cite{BFMR2022prx}, most notably, in the form of nonreciprocal Cahn-Hilliard models \cite{YoBM2020pnasusa,SaAG2020prx,FrWT2021pre}. The latter add nonreciprocal interactions to classical Cahn-Hilliard models \cite{Cahn1965jcp} (model-B in \cite{HoHa1977rmp}) that describe the dynamics of phase-separation,  e.g., in binary or ternary mixtures \cite{NaHe2001ces,MKHK2019sm}. In particular, the resulting nonreciprocal Cahn-Hilliard models represent two conservation laws with nonvariational coupling. It is shown that this coupling may result in coarsening traveling and oscillatory states \cite{YoBM2020pnasusa,SaAG2020prx,FrWT2021pre}, arrest or suppression of coarsening \cite{FrWT2021pre}, formation of small-scale spatial (Turing) patterns as well as stationary, traveling and oscillatory localized states \cite{FrTh2021ijam} - all features that are forbidden in standard reciprocal Cahn-Hilliard models.

However, these nonreciprocal Cahn-Hilliard models are introduced on phenomenological grounds by symmetry considerations, but no derivation of the field theory from a microscopic description or other deeper justification has been provided yet. Here, we show that the model indeed merits extensive study as it actually represents one of the universal equations of pattern formation. One may even argue that it corresponds to a
``missing amplitude equation'' for the basic eight types of linear instabilities in spatially extended isotropic homogeneous systems that can be described by scalar fields. An amplitude (or envelope) equation describes the universal bifurcation behavior characterizing the spatio-temporal dynamics in the vicinity of the threshold of a single instability or of several simultaneous instabilities, and can be systematically derived in a weakly nonlinear approach \cite{Hoyle2006}. The mentioned eight instability types result from the combination of three features: (i) large-scale vs.\ small-scale instability, (ii) stationary vs.\ oscillatory instability, and (iii) instability without and with conservation law(s). The spatial and temporal character of an instability encoded in features (i) and (ii) is well captured in the classification of instabilities by Cross and Hohenberg \cite{CrHo1993rmp}, and the four corresponding amplitude equations for systems without conservation law are very well studied. One example is the complex Ginzburg-Landau equation \cite{ArKr2002rmp} valid near the onset of a large-scale oscillatory (aka Hopf or type~III$_\mathrm{o}$ \cite{CrHo1993rmp}) instability. An overview of the basic eight instability types in our amended classification, their dispersion relations and 
seven existing amplitude equations is provided in Section~1 
 of the Supplementary Material.

However, the consequences of conservation laws in the full range of pattern-forming systems are less well studied: Small-scale stationary and oscillatory cases with a conservation law are considered in \cite{MaCo2000n} and \cite{WiMC2005n}, respectively, with applications to pattern formation in the actin cortex of motile cells \cite{BeGY2020c,YoFB2022prl}, in crystallization \cite{TARG2013pre}, and in magnetoconvection \cite{Knob2016ijam}. However, only recently it was shown that the standard single-species Cahn-Hilliard equation does not only describe phase separation in a binary mixture \cite{Cahn1965jcp,Bray1994ap} but furthermore can be derived as an amplitude equation valid in the vicinity of a large-scale stationary instability in a system with a single conservation law \cite{BeRZ2018pre}. In consequence, close to onset, a reaction-diffusion system with one conservation law as, e.g., discussed in \cite{IsOM2007pre,OICK2007pcb,EiIM2012dcdsb,WeBK2014if,BDGY2017nc,CREW2018pcb,HaFr2018np,BrHF2020prx,BeGY2020c,BWHY2021prl}, can be quantitatively mapped onto a Cahn-Hilliard equation. Similarly, the equation captures core features of certain collective behavior in chemotactic systems \cite{RaZi2019pre} and for cell polarization in eukaryotic cells \cite{BeZi2019po}.

This leaves only one of the eight cases unaccounted for, namely, the large-scale oscillatory instability with conservation laws, that we call \textit{conserved-Hopf instability}.  In the following, we consider active systems with two conservation laws and show that the general nonreciprocal Cahn-Hilliard model emerges as a corresponding universal amplitude equation. Thereby, all the particular phenomenological models studied in \cite{YoBM2020pnasusa,SaAG2020prx,FrWT2021pre,FrTh2021ijam} are recovered as special cases. This also applies to the complex Cahn-Hilliard equation appearing as a mass-conserving limiting case in Ref.~\cite{Zimm1997pa}.

Before we embark on a general derivation of the amplitude equation we emphasize its applicability to the wide spectrum of systems where the conserved-Hopf instability and related intricate nonlinear oscillatory behavior can occur: A prominent example is the spatio-temporal pattern formation of proteins vital for cellular processes. Although chemical reactions cause conformation changes of proteins, their overall number is conserved on the relevant time scale, e.g., MinE and MinD in ATP-driven cellular Min oscillations \cite{HaFr2018np}. Such an instability can also be expected in other reaction-diffusion systems with more than one conservation law, e.g., the full cell polarity model in Ref.~\cite{OICK2007pcb}. Relevant examples beyond reaction-diffusion systems include oscillations in two-species chemotactic systems \cite{Wola2002ejam}, an active poroelastic model for mechanochemical waves in cytoskeleton and cytosol \cite{RAEB2013prl}, thin liquid layers covered by self-propelled surfactant particles \cite{PoTS2016epje},  oscillatory coupled lipid and protein dynamics in cell membranes \cite{JoBa2005pb}, multi-component phase-separating reactive or surface-active systems \cite{OSIK2020pre,WSH2021jcp}, and two-layer liquid films or drops on a liquid layer with mass transfer \cite{CSS2017l} or heating \cite{PBMT2005jcp} where the two interfaces may show intricate spatio-temporal oscillation patterns \cite{CSS2017l,NeSi2017pf,NeSh2016jpat}. 

In most cases, the two conserved quantities correspond to concentration fields, film or drop thickness profiles, particle number densities and the conserved-Hopf instability occurs as primary instability. However, another class of examples exists where it appears as a secondary instability. For example, in Marangoni convection the interaction between a large-scale deformational and a small-scale convective instability is described by coupled kinetic equations for the film height and a complex amplitude \cite{GNPR1997pd}. There, liquid layer profile and phase of the complex amplitude represent the two conserved quantities and the occurring conserved-Hopf instability corresponds to an oscillatory sideband instability.

Systems like the given examples that feature two conservation laws and exist in a sustained out-of-equilibrium setting can become unstable through a conserved-Hopf instability, i.e., the linear marginal mode (growth rate $\Delta(k_\mathrm{c})=0$) occurs at zero wavenumber ($k_\mathrm{c}=0$) and zero frequency ($\Omega(k_c)=\Omega_\mathrm{c}=0$) but is oscillatory arbitrarily close to onset. This is determined via a linear stability analysis of the trivial uniform steady state yielding the dispersion relations $\lambda_\pm(k)$ of the dominant pair of complex conjugate modes where $\Delta=\Re\lambda_\pm$ and $\Omega=\pm\Im\lambda_\pm$. Although $\lambda_\pm(k=0)=0$ always holds as the two conservation laws imply the existence of two neutral modes, the mode is nevertheless oscillatory at arbitrarily small wavenumbers. In other words, directly beyond instability onset the system undergoes large-scale small-frequency oscillations, i.e., the conservation laws imply that the first excited mode has the smallest wavenumber compatible with the domain boundaries and oscillates on a correspondingly large time scale as $\Omega\to0$ for $k\to 0$. In consequence, the weakly nonlinear behavior is not covered by any of the seven amplitude equations summarized in Section~1 
 of the Supplementary Material.

\begin{figure}[tbh]
	\includegraphics[width=0.8\columnwidth]{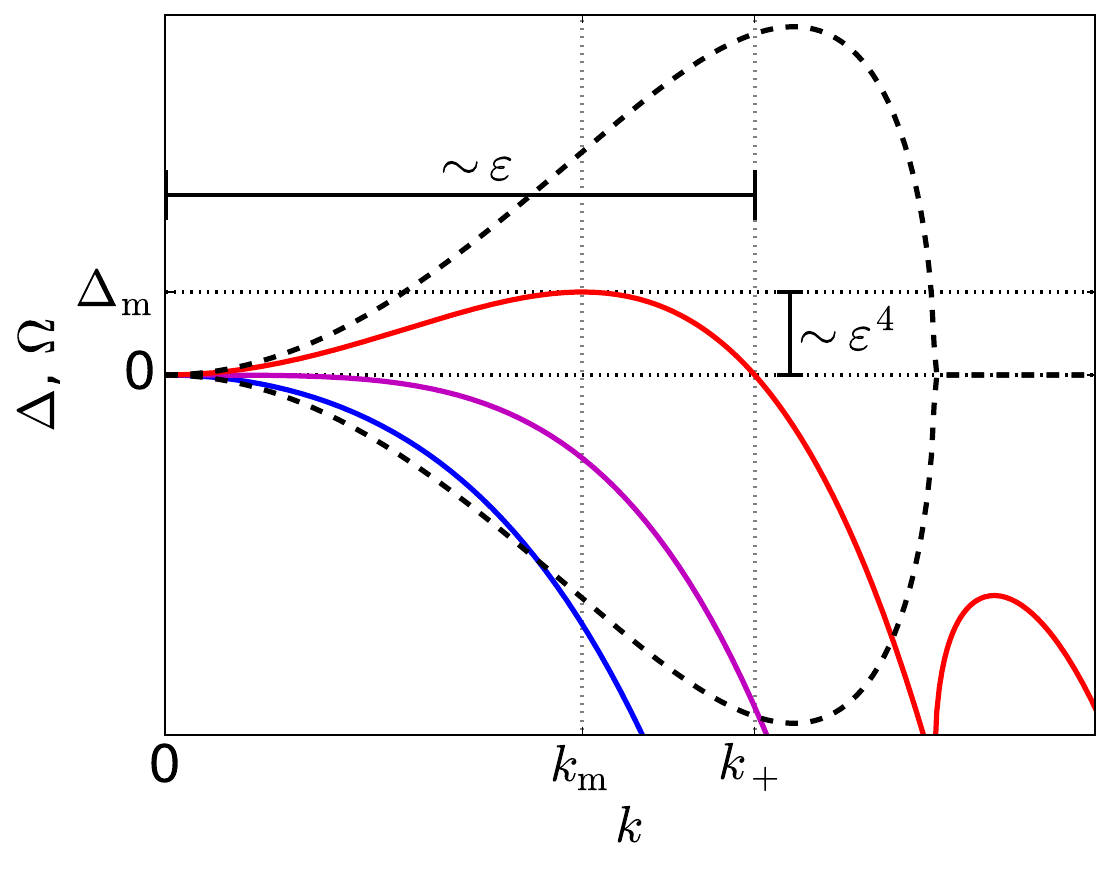}
	\caption{Linear growth rates $\Delta(k)=\Re\lambda_\pm(k)$ in dependence of the wavenumber $k$ below (solid blue line, $\delta<0$), at (solid purple line, $\delta=0$) and above (solid red line, $\delta>0$) the threshold of a conserved-Hopf instability as described by the dispersion relation $\lambda_\pm(k)$ given by the series expansion Eq.~\eqref{eq:series}. The black dashed lines give the frequencies $\pm\Omega(k)=\Im\lambda_\pm(k)$ that are identical in all three cases. Labeled thin dotted lines and solid bars indicate typical quantities and scalings above onset as described in the main text.
	}
	\label{fig:onset_hopfconserved}
\end{figure}

Dispersion relations below, at and above the threshold of a conserved-Hopf instability are sketched in Fig.~\ref{fig:onset_hopfconserved} and are at small $k$ given by
\begin{align}
	\lambda_\pm(k) =& \Delta(k) \pm \mathrm{i} \Omega(k)~\nonumber\\ 
  \mathrm{with}\,\, \Delta(k) =& \delta k^2 - \tilde \delta k^4 + \mathcal{O}(k^6) \label{eq:series}\\
  \mathrm{and}\quad \Omega(k) =& \omega k^2 +\tilde \omega k^4 + \mathcal{O}(k^6).~\nonumber
\end{align}
The onset occurs when $\delta$ becomes positive while $\tilde \delta>0$. Above onset, Eq.~\eqref{eq:series} indicates a band of wavenumbers $0<k<k_+=\sqrt{\delta/\tilde \delta}$ that exponentially grow. The fastest mode is at $k_{\text{m}}= k_+/\sqrt{2}$ and has the growth rate $\Delta_\text{m}=\delta^2/(4\tilde \delta)$.

To determine an amplitude equation that captures the bifurcation structure characterizing the spatio-temporal pattern formation in the vicinity of the onset of a conserved-Hopf bifurcation with a dispersion relation as depicted in Fig.~\ref{fig:onset_hopfconserved} we apply a weakly nonlinear approach \cite{Hoyle2006}. First, we introduce a smallness parameter $\varepsilon$ with $|\varepsilon| \ll 1$ and consider the system close to onset where $\delta= \delta_2 \varepsilon^2$. From hereon subscript numerals indicate the order in $\varepsilon$ of the corresponding term. Then, the width of the band of growing wavenumbers and the maximal growth rate scale as $\varepsilon$ and $\varepsilon^4$, respectively. This determines the additional large spatial scale $\vec X= \varepsilon \vec x$ and slow timescale $T=\varepsilon^4 t$ relevant for the dynamics. Additionally, Eq.~\eqref{eq:series} indicates that the leading order oscillation frequency scales like $\Omega\approx \omega k^2 \sim \varepsilon^2$. This implies that a second slow timescale $\tau=\varepsilon^2 t$ has to be considered.

Specifically, we now consider a general homogeneous isotropic multi-component system with two conservation laws
\begin{align}\label{eq:2Conssystem_main}
\begin{split}
\partial_t \rho =& -\vec \nabla\cdot\left( Q(\vecg u) \vec \nabla \eta(\vecg u, \vec \nabla) \right)~\\
\partial_t \sigma =& -\vec \nabla\cdot \left( R(\vecg u) \vec \nabla \mu(\vecg u, \vec \nabla) \right)~\\
\partial_t \vecg{n} =& \vecg{F}(\vecg u, \vec \nabla),
\end{split}
\end{align}
i.e., coupled kinetic equations for two conserved ($\rho$ and $\sigma$) and $N$ nonconserved [$\vecg{n}=(n_1,\dots,n_N)$] scalar field variables. Note that Eq.~\eqref{eq:2Conssystem_main} can represent any of the examples mentioned above. From here onwards, $\vecg u =\left(\rho, \sigma, \vecg{n}\right)$ is used as abbreviation where convenient. The dynamics of the two conserved quantities is given by the divergence of corresponding fluxes that consist of the product of a mobility ($Q$ or $R$) and the gradient of a nonequilibrium (chemical) potential ($\eta$ or $\mu$) that, in general, still depends on spatial derivatives $\vec \nabla$. The dynamics of the nonconserved quantities is given by the vector $\vecg{F}$ of general functions of fields and their spatial derivatives. In 
the simplest case, the system may represent a reaction diffusion system with $N+2$ species that has been rearranged (similar to \cite{BrHF2020prx})\footnote{For reaction-diffusion systems with mass conservation, quantities like $\eta$ and $\mu$ correspond to linear combinations of concentration fields called ``mass-redistribution potentials'' in Ref.~\cite{BrHF2020prx,BWHY2021prl}.} to explicitly show the two conservation laws. More complicated examples include multi-field thin-film descriptions where two components are conserved \cite{NeSh2016jpat} and multi-species membrane models showing phase separation and chemical reactions \cite{JoBa2005pb}.
For an active system the potentials can not be obtained as variational derivatives of a single underlying energy functional. Here, we sketch the derivation of an
amplitude equation for the conserved-Hopf instability (Fig.~\ref{fig:onset_hopfconserved}) of a homogeneous steady state of a general system~\eqref{eq:2Conssystem_main} while Section~2 
 of the Supplementary Material present the complete algebra.

To perform the weakly nonlinear analysis valid in the vicinity of instability onset, we expand all fields in $\varepsilon$, i.e.,
$\vecg u  (\vec{X},\tau,T)= \vecg u_0 + \varepsilon \vecg u_1(\vec{X},\tau,T) + \varepsilon^2 \vecg u_2(\vec{X},\tau,T) + \dots$, where $\vecg u_0$ is the steady uniform state with $\vecg F(\vecg u_0)=0$ and $\vecg u_i(\vec{X},\tau,T)$, $i=1,2,\dots$ are the deviations that describe the (weakly) nonlinear behavior. We take account of the above discussed scaling of space and time implied by the dispersion relation by writing $\vec \nabla_{\!\vec x} = \varepsilon \vec \nabla_{\!\vec X}$ and $\partial_t= \varepsilon^2 \partial_\tau + \varepsilon^4 \partial_T$, respectively.
With this we can then consider Eqs.~\eqref{eq:2Conssystem_main} order by order.
The scaling implies that we need to successively consider all orders up to $\mathcal{O}(\varepsilon^5)$ to discover evolution equations that capture dynamic effects on the slow timescale $T$.

In principle, at each order we first determine the nonconserved fields as (nonlinear) functions of the conserved fields, reflecting that the dynamics of the former is slaved to the latter. Second, we obtain the continuity equations to the corresponding order by inserting the obtained expressions into the appropriate mobilities and potentials similar to Taylor-expanding them. In particular, at order $\varepsilon$, the contributions of the two continuity equations vanish and the remaining $N$ equations become a homogeneous linear algebraic system for the slaved quantities, solved by $\vecg n_1(\vec{X},\tau,T) = \vecg n_\rho \rho_1(\vec{X},\tau,T) + \vecg n_\sigma \sigma_1(\vec{X},\tau,T)$ where $\vecg n_\rho$ and $\vecg n_\sigma$ correspond to the zero eigenmodes $(1,0,\vecg n_\rho)$ and $(0,1,\vecg n_\sigma)$ of the dominant eigenspace at $k=0$.\footnote{For reaction-diffusion systems the reconstructed slaved $\vecg n$ correspond to the ``local reactive equilibria'' introduced in \cite{HaFr2018np}. This identification will allow one to apply many aspects of the phase-space methods advanced in \cite{HaFr2018np,BrHF2020prx,BWHY2021prl} to the wider class of systems considered here.}

At order $\varepsilon^2$, again the continuity equations are trivially fulfilled, and the remaining equations form an inhomogeneous linear algebraic system for the $\vecg n_2$. Thereby, the inhomogeneity is nonlinear in lower order quantities.
At order $\varepsilon^3$, the first nonvanishing contributions from the continuity equations appear, that, after eliminating $\vecg n_1$, correspond to linear equations for 
$\rho_1$ and $\sigma_1$. They provide the conditions 
for the instability onset at $\delta=0$ in Eq.~\eqref{eq:series}. They also capture the leading order oscillations with frequency $\omega$ on the time scale $\tau$ by an antisymmetric dynamic coupling that represents a nonreciprocal coupling of lowest order (a structure equivalent to the Schr\"odinger equation for a free particle). Also for the $\vecg n_3$ an inhomogeneous linear algebraic system emerges. At the subsequent order $\varepsilon^4$, further contributions to the evolution on the time scale $\tau$ are obtained from the continuity equations. Finally, at order $\varepsilon^5$ we obtain expressions for $\partial_T \rho_1$ and $\partial_T \sigma_1$. Using the earlier obtained results for $\vecg n_1$, $\vecg n_2$, and $\vecg n_3$, the complete continuity equations at this order can be written as nonlinear functions of the $\rho_i$ and $\sigma_i$. This provides the weakly nonlinear expression for the leading order time evolution on the timescale $T$. 
Next, the expressions found at the different orders are recombined, in passing ``inverting'' the scalings and expansions of time, coordinates and fields $\rho$ and $\sigma$. The resulting amplitude equation corresponds to a generalized nonreciprocal Cahn-Hilliard model (i.e., two nonreciprocally coupled Cahn-Hilliard equations) and is given in Section~2 of the Supplementary Material. In the common case of constant mobilities ($Q=Q_0$ and $R=R_0$) in Eq.~\eqref{eq:2Conssystem_main}, cross-couplings in the highest-order derivatives may be removed by a principal axis transformation, resulting in
  \begin{align}	 \label{eq:general_nonCH_Amain}
\partial_t A &= \vec\nabla^2 \left[\alpha_1 A +  \alpha_2 B + N_A(A,B) - D_A \vec\nabla^2 A \right]\nonumber\\
\partial_t B &= \vec\nabla^2 \left[\beta_1 A +  \beta_2 B + N_B(A,B) - D_B \vec\nabla^2 B \right]
  \end{align}
where the spatially slowly varying real amplitudes $A$ and $B$ are linear combinations of the deviations of the conserved fields from their mean values, $D_A$ and $D_B$ are effective interface rigidities, and $N_A$ and $N_B$ are general cubic  polynomials in $A$ and $B$, e.g., $N_A=\alpha_3 A^2 + \alpha_4 A B + \alpha_5 B^2 + \alpha_6 A^3 +  \alpha_7 A^2 B +  \alpha_8 A B^2 + \alpha_9 B^3$. All parameters are real.\footnote{Rescaling time, space, and the two fields one may set four parameters to unity, and affine transformations in the fields can be employed to eliminate a few terms. System-intrinsic symmetries would further simplify the cubic polynomials. For instance, an inversion symmetry in the conserved fields may eliminate all quadratic terms.}

The derived general nonreciprocal two-component Cahn-Hilliard model describes the universal bifurcation behavior in the vicinity of any conserved-Hopf instability independently of the particular system studied -- all such systems and most of their parameters at instability onset are encoded in the rich parameter set of the derived equations. It should further be noted that the derived general model encompasses further primary bifurcations as it actually corresponds to the amplitude equation for an instability of higher codimension. This is shown in Section~3 of the Supplementary Material at the example of a Cahn-Hilliard instability of codimension two. In other words, the derived amplitude equation may be considered as belonging to a higher level of a hierarchy of such equations. It captures several qualitatively different linear instability scenarios.
Such hierarchies are useful to understand the qualitative differences and transitions between instability types. Amplitude equations on a higher hierarchy level describe the bifurcation behavior close to higher codimension points, i.e., the behavior in the vicinity of several different instabilities. In the limiting case where only one of the contained  instabilities is close to its onset, the higher level equation can often be reduced to a simpler lower level equation~\cite{RoDo1998pd}. However, such a further reduction of the derived nonreciprocal Cahn-Hilliard equation remains a task for the future.

Note that the presence of additional subdominant neutral modes (e.g., resulting from additional conservation laws) or the simultaneous onset of several distinct instabilities would (possibly in extension of the present work) also result in amplitude equations on a higher level of the ``codimension hierarchy'' \cite{DLDB1996pre,YoFB2022prl}.

It is an interesting observation that the various \textit{ad-hoc} nonreciprocal Cahn-Hilliard models studied in \cite{YoBM2020pnasusa,SaAG2020prx,FrWT2021pre} emerge as special cases of the equation derived here.\footnote{In contrast to part of the \textit{ad-hoc} examples the here derived general model is generic in the sense of Ref.~\cite{FHKG2023pre}.} Table~3 in Section~2 
 of the Supplementary Material provides the corresponding parameter choices in Eq.~\eqref{eq:general_nonCH_Amain}. Two other limiting cases are also included:  (i) if certain symmetries between coefficients hold, one may introduce a complex amplitude $C= A + \mathrm{i} B$ and present Eq.~\eqref{eq:general_nonCH_Amain} as a complex Cahn-Hilliard equation $\partial_t C =  -G\vec \nabla^2 \left[ \varepsilon + (1 + \mathrm{i}b) \vec\nabla^2 - (1 + \mathrm{i} c) |C|^2 \right]C$, i.e., as a complex Ginzburg-Landau equation with an additional outer Laplace operator reflecting the conservation property, as briefly considered in Ref.~\cite{Zimm1997pa}. This, in passing clarifies that Eq.~\eqref{eq:general_nonCH_Amain} is more than just a ``conserved complex Ginzburg-Landau equation'' because it does not show its phase-shift invariance. (ii) Imposing another symmetry between coefficients renders the coupled equations variational. Then they represent a generic model for the dynamics of phase separation in a ternary system \cite{MKHK2019sm,Ma2000jpsj}.

To conclude, we have derived an amplitude equation valid in the vicinity of a conserved-Hopf bifurcation and as well at related bifurcations of higher codimension. It qualitatively captures transitions generically occurring in the wide variety of out-of-equilibrium systems that feature two conservation laws. Note that close to the conserved-Hopf instability it also provides a rather good quantitative description of the bifurcation structure. This is exemplified in Section~4 of the Supplementary Material~\footnote{Note that the Supplementary Material additionally contains Refs.~\cite{Knob1992,ThAP2016prf,JPMW2014jem,DWCD2014ccp,UeWR2014nmma}.} where the amplitude equation is derived and analyzed in comparison with the full system for the relatively simple case of a three-component reaction-diffusion system with two conservation laws. As the latter reduce the local phase space (defined as in Ref.~\cite{BrHF2020prx}) to one dimension, the emerging behavior will be much less complex than seen in the Min system \cite{HaFr2018np} and other high dimensional cases \cite{PoTS2016epje}.

The derived equation forms part of the hierarchy of universal amplitude equations for the above discussed eight basic instabilities. Thus, its relevance for the classification of pattern forming behavior close to the onset of instabilities resembles that of the complex Ginzburg-Landau equation that describes the universal bifurcation behavior in the vicinity of a standard Hopf instability in systems without conservation laws \cite{CrHo1993rmp,ArKr2002rmp,Hoyle2006,Pismen2006}. However, one has to add restrictively that the large number of parameters of the derived generic model might limit its practical use as a complete parametric study of all generic behaviors is prohibitively costly. Still its study has already started to form a valuable bridge between the analysis of the many specific models and the set of amplitude equations on a lower hierarchy level (that still needs completion). In cases where the primary bifurcation is subcritical (e.g., for the Min oscillations \cite{HaFr2018np}), even higher order amplitude equations might be insufficient to faithfully predict the spatio-temporal behavior. Then weakly and fully nonlinear approaches should be employed in a complementary manner.

Although it is known that the conserved-Hopf instability is related to phenomena that are not covered by the complex Ginzburg-Landau equation \cite{NeSh2016jpat} only very few studies have considered its (weakly) nonlinear behavior by corresponding amplitude equations~\cite{GNPR1997pd,OrNe2004pre}, normally, in special cases. On the one hand, Ref.~\cite{GNPR1997pd} restricts its focus to amplitude equations for spatially periodic traveling and standing waves, and on the other hand, 
Ref.~\cite{OrNe2004pre} deals with a particular case without reflection symmetry where one of the two conservation laws is weakly broken. The universal character of the model derived here, implies that literature results on the onset of motion and oscillations \cite{YoBM2020pnasusa,SaAG2020prx,FrWT2021pre} and as well potentially on the suppression of coarsening and the existence of localized states \cite{FrWT2021pre,FrTh2021ijam} may be applied to the class of out-of-equilibrium systems that undergo a conserved-Hopf instability. In consequence, spatio-temporal patterns occurring in a wide spectrum of systems from protein dynamics within cells and on membranes \cite{JoBa2005pb,HaFr2018np}, chemotactic systems of organisms \cite{Wola2002ejam}, coupled cytoskeleton and cytosol dynamics \cite{RAEB2013prl}, multi-component phase-separating reactive, surface-active or active systems \cite{OSIK2020pre,WSH2021jcp,PoTS2016epje}, to two-layer liquid films with heating or mass transfer \cite{PBMT2005jcp, CSS2017l,NeSi2017pf} should be further studied to identify their common universal features as out-of-equilibrium systems with conservation laws as well as characterizing differences that may prompt a further development of the hierarchy of amplitude equations. 

However, the present work has entirely focused on isotropic homogeneous systems described by scalar fields, implying that systems like the active Ising model in \cite{SoTa2013prl,SoTa2015pre} are not covered as they involve a pseudo-scalar. The dispersion relations of such systems with conservation laws have properties different from the ones considered here. It would be highly interesting to produce a systematics similar to the one proposed here for systems involving pseudo scalars. To our knowledge, so far only a few cases have been treated by weakly nonlinear theory.

\textit{Acknowledgement}
TFH and UT acknowledge support from the doctoral school ‘‘Active living fluids’’ funded by the German-French University (Grant No. CDFA-01-14). In addition, TFH thanks the foundation ``Studienstiftung des deutschen Volkes'' for financial support.
The authors thank Daniel Greve for fruitful discussions and the anonymous reviewers for constructive comments.

\clearpage
\onecolumngrid
\appendix

\section*{Supplementary Material}

\subsection{Dispersion relations and amplitude equations}
\label{sec:app:disp-amp}
This section of the Supplementary Material gives an overview of the eight types of linear instabilities of uniform steady states occurring in homogeneous isotropic systems described by scalar field variables. In particular, we discuss the corresponding dispersion relations and the seven well investigated amplitude equations (also called envelope equations) obtained by weakly nonlinear theory in the vicinity of the instability thresholds.

\begin{figure}
\includegraphics[height=0.27\hsize]{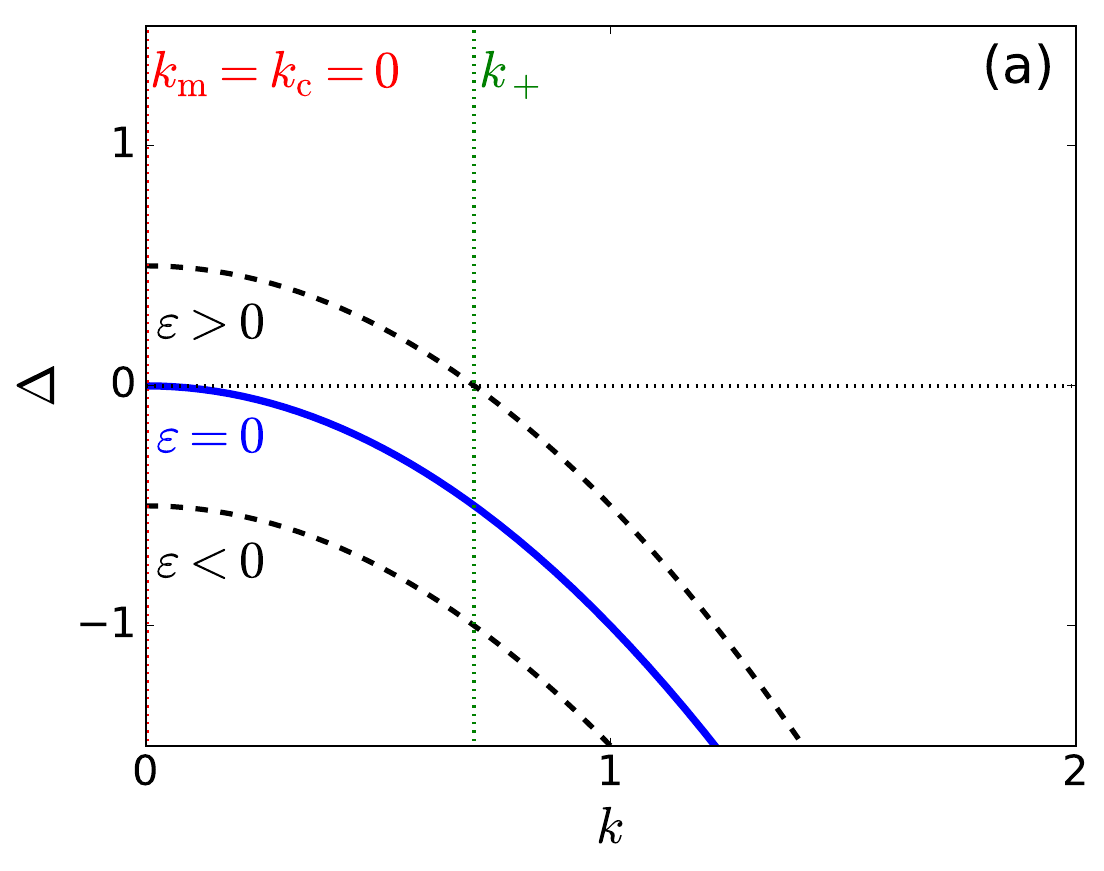}
\hspace{0.05\hsize}
\includegraphics[height=0.27\hsize]{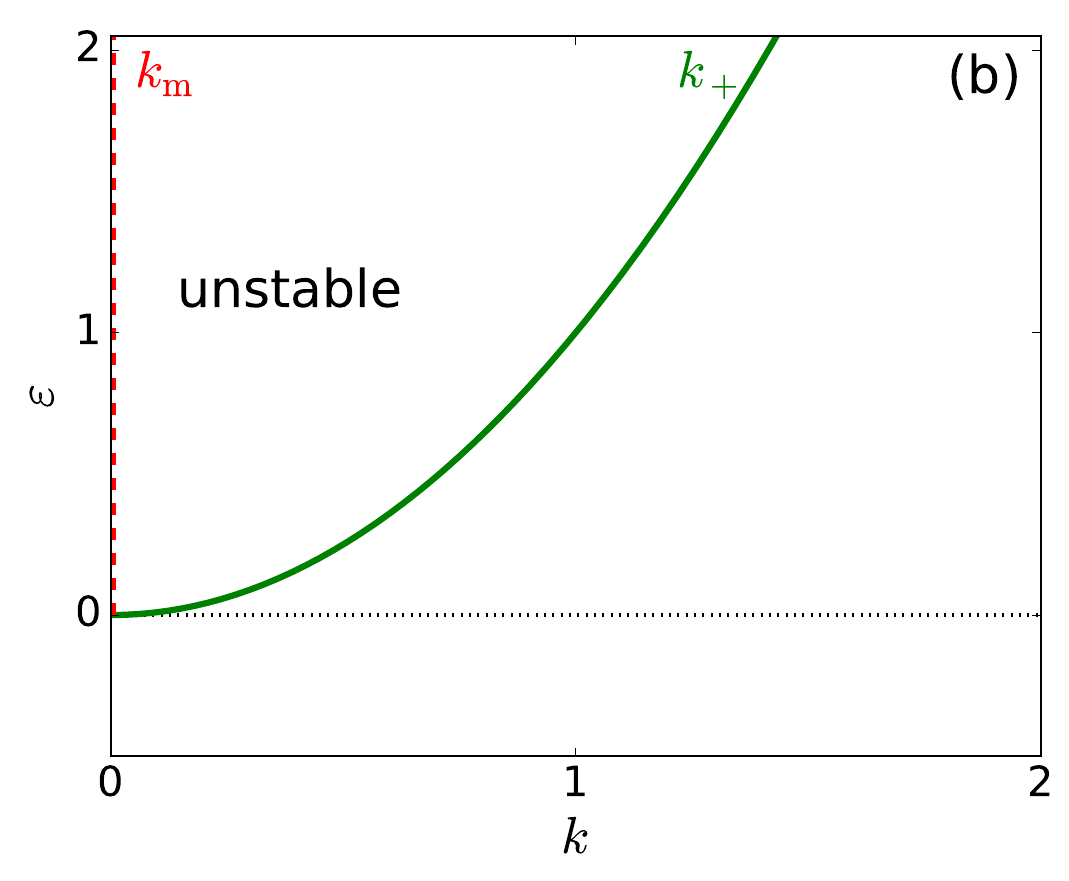}\\
\includegraphics[height=0.27\hsize]{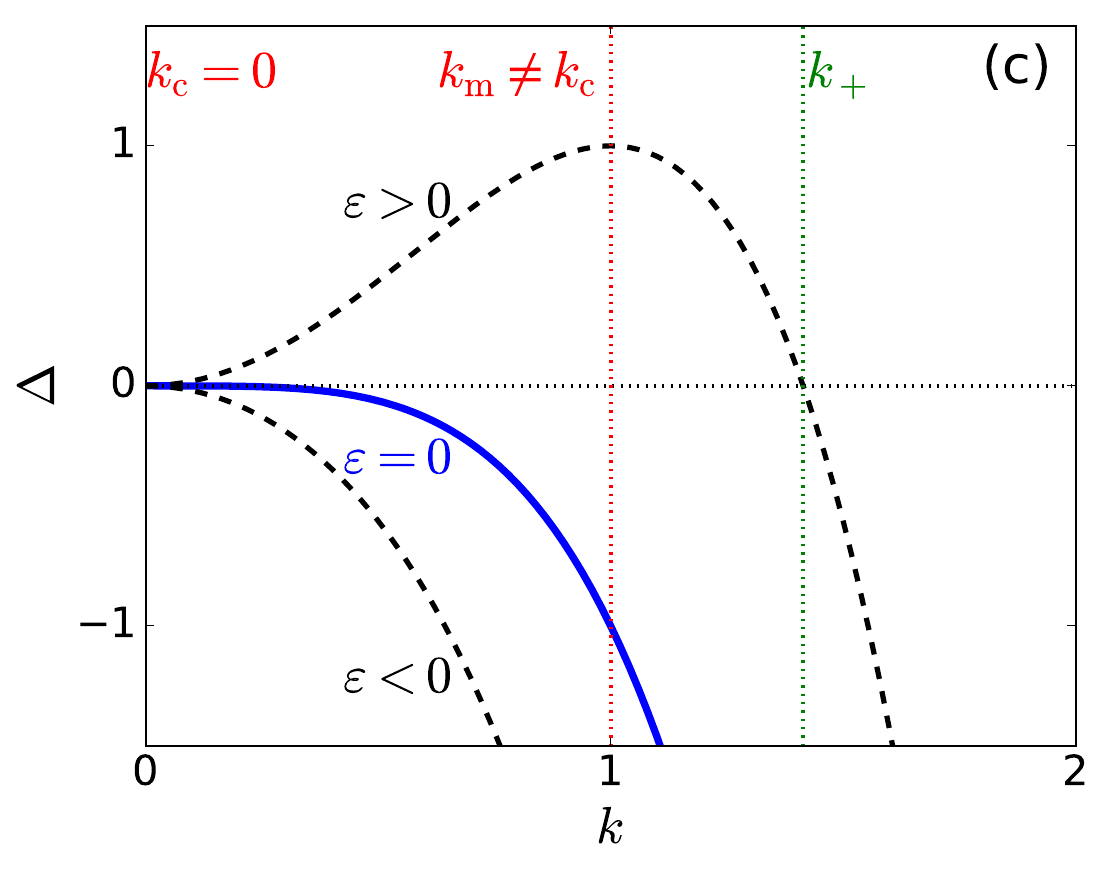}
\hspace{0.05\hsize}
\includegraphics[height=0.27\hsize]{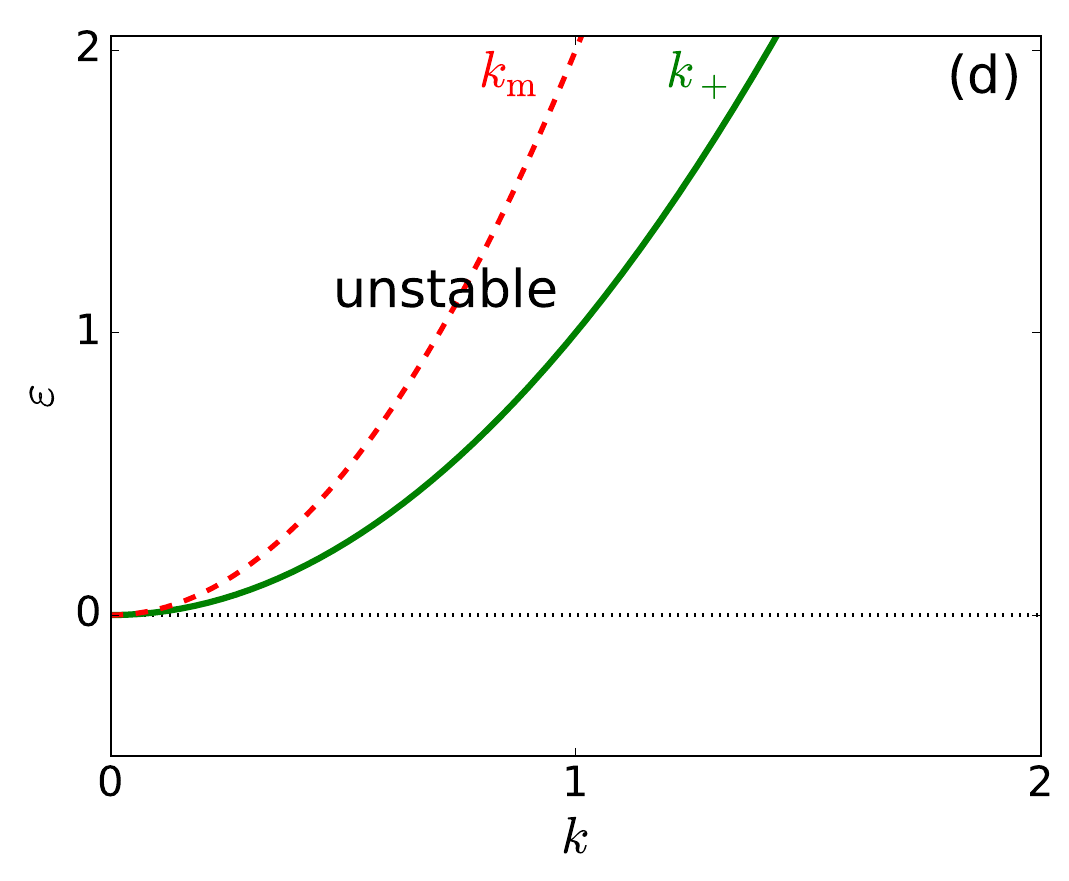}\\
\includegraphics[height=0.27\hsize]{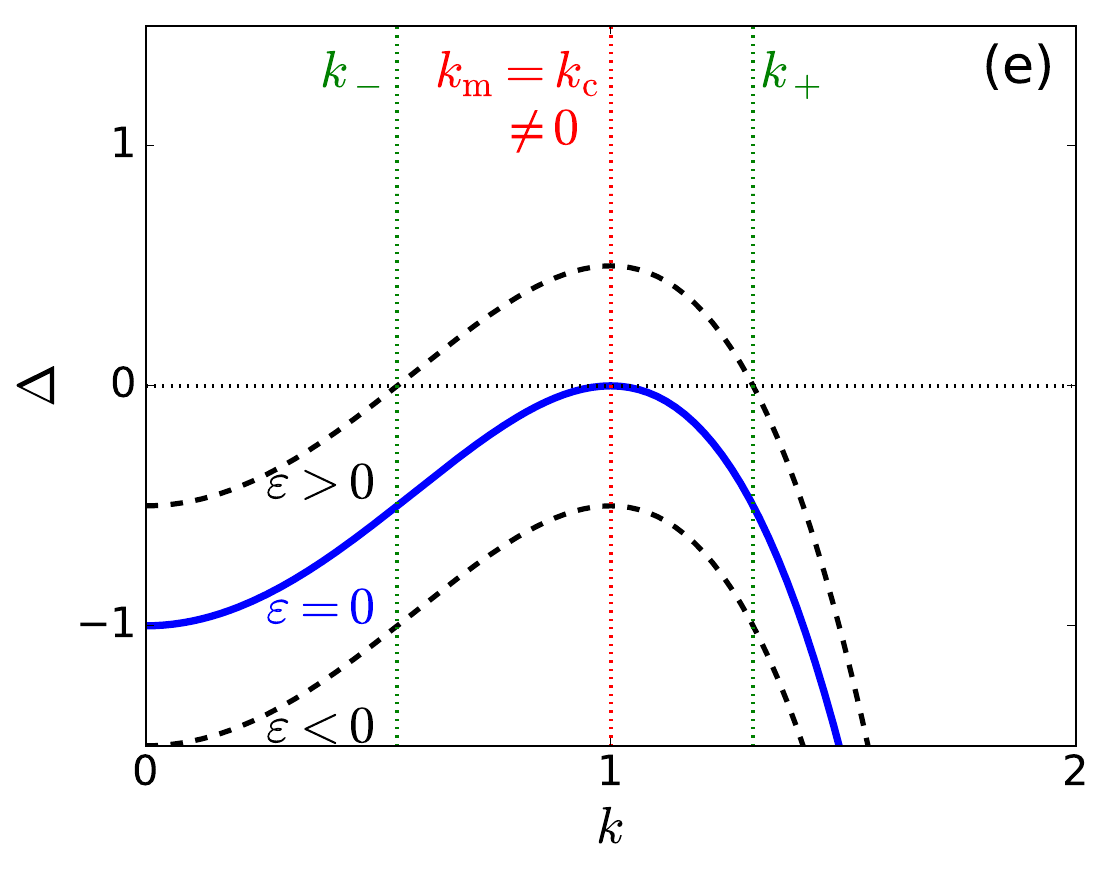}
\hspace{0.05\hsize}
\includegraphics[height=0.27\hsize]{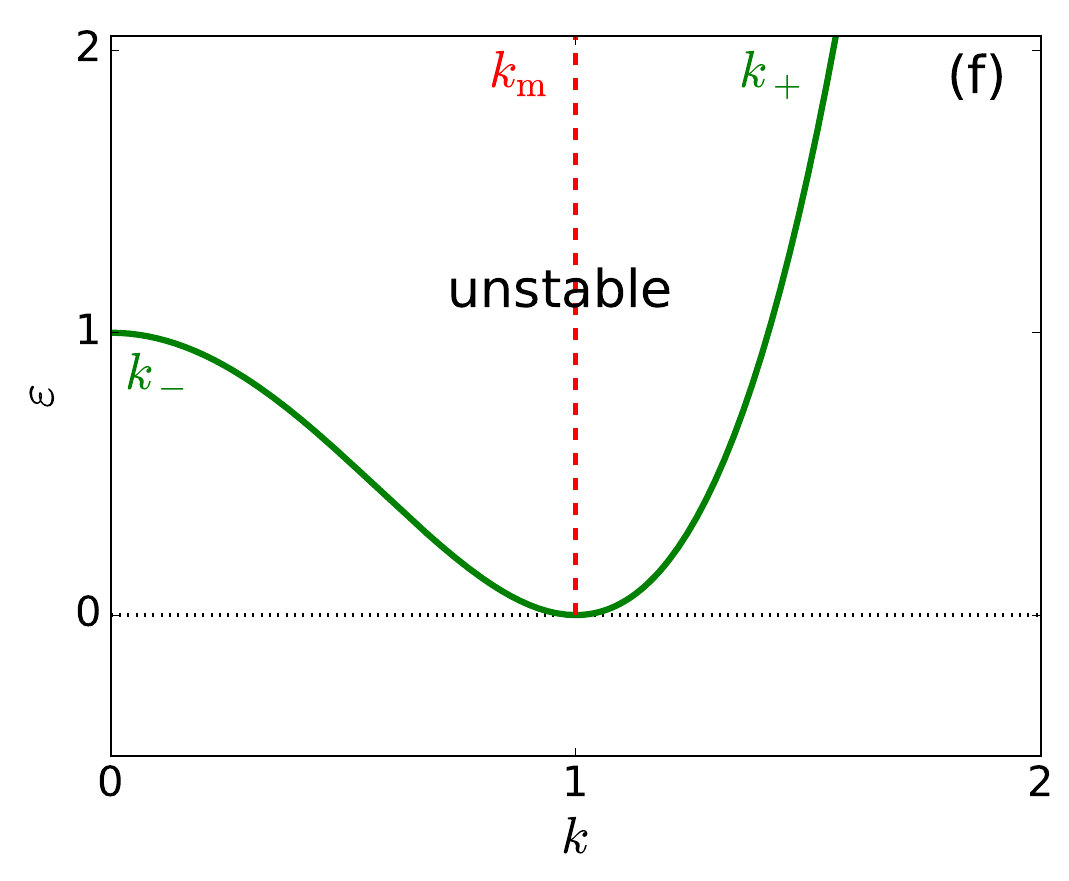}\\
\includegraphics[height=0.27\hsize]{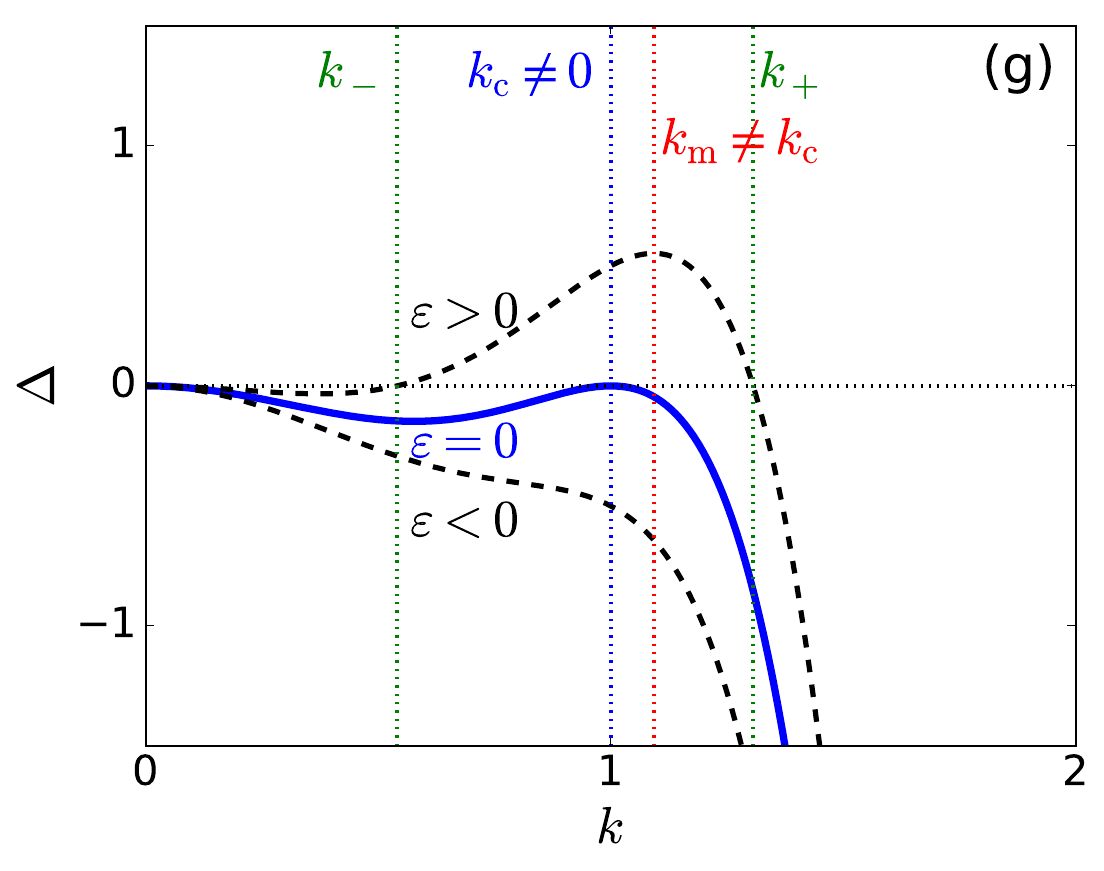}
\hspace{0.05\hsize}
\includegraphics[height=0.27\hsize]{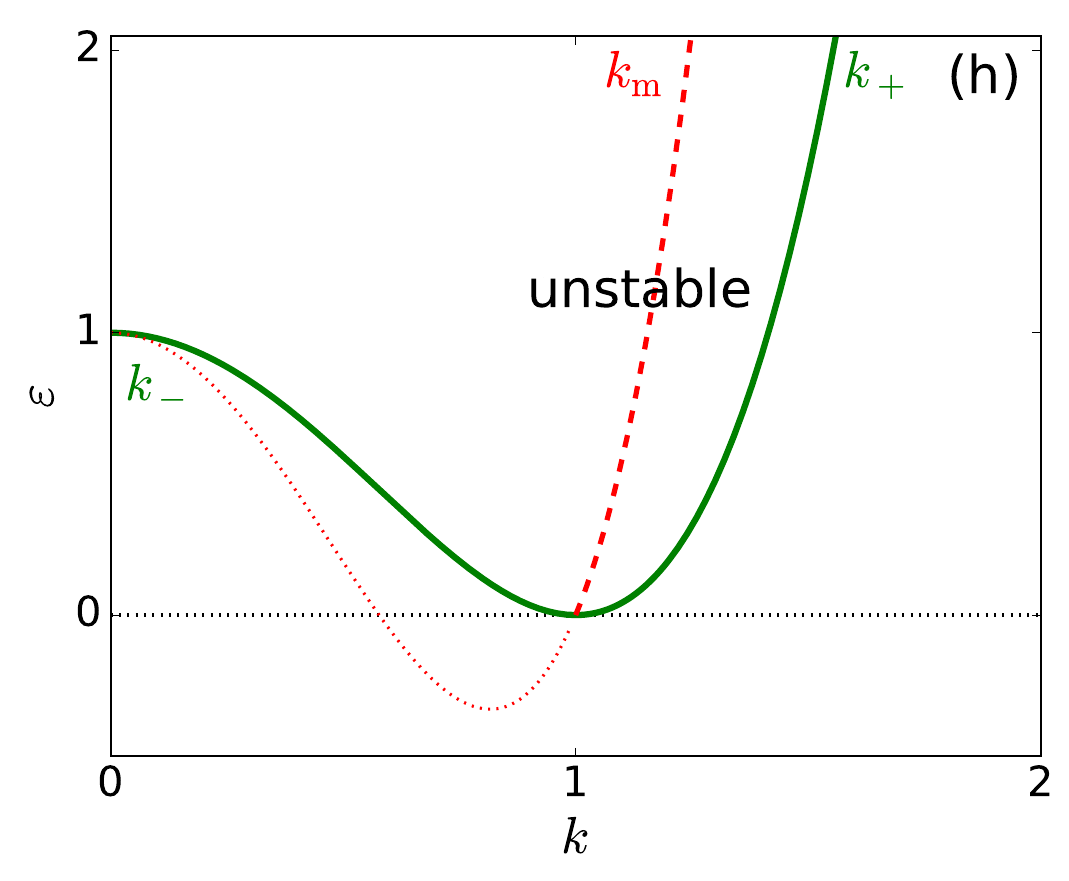}\\
\caption{\small Classification of dispersion relations distinguishing large-scale and small-scale instabilities as well as instabilities without and with conservation law(s). Shown are (left) the real part $\Delta=\mathrm{Re}\,\lambda(k)$ of the dispersion relation and (right) the position of marginal wavenumber(s) $k_\pm$ (solid line) and fastest growing wavenumber $k_\mathrm{m}$ (dashed line) in the plane spanned by wavenumber $k$ and control parameter $\varepsilon$. For each of the four cases there exist stationary and oscillatory variants. In the left panels we also indicate the critical wavenumber $k_\mathrm{c}$ where the instability onset occurs (at $\varepsilon=0$, heavy solid blue line). Dashed lines give $\Delta(k)$ below ($\varepsilon<0$) and above ($\varepsilon>0$) onset. Shown are (a,b) the nonconserved homogeneous (Allen-Cahn and Hopf) instability, (c,d) the conserved large-scale (Cahn-Hilliard and conserved-Hopf)  instability, (e,f) the nonconserved small-scale (Turing and wave) instability, and (g,h) the conserved small-scale (conserved-Turing and conserved-wave) instability. Naming conventions are summarized in Table~\ref{tab:instab}. \label{fig:linstab}}
  \end{figure}

The restriction to homogeneous isotropic systems implies that underlying model equations are translation- and rotation-invariant. For simplicity, in the following we only consider spatially one-dimensional systems, i.e., isotropy becomes reflection symmetry (also called parity symmetry). The considered multi-component order parameter field $\vecg u(x,t)$ represents a set of scalars. All occurring conservation laws are assumed to be local, i.e., the kinetic equation(s) for corresponding conserved fields $\rho$ have the form of a continuity equation $\partial_t \rho = -\partial_x j$ where $j(x,t)$ is a flux that may depend linearly or nonlinearly on all components of $\vecg u(x,t)$ and their spatial derivatives.

Parity symmetry implies that each r.h.s.\ term features an even number of spatial derivatives.  Here, $x$ and $t$ are position and time while $\partial_x$ and $\partial_t$ are the corresponding partial derivatives.

  The linear stability of steady uniform states $\vecg u(x,t)=\vecg u_0$ is determined by adding small-amplitude perturbations
    that respect the boundary conditions and show an exponential time dependence $e^{\lambda t}$ where $\lambda$ is a real or complex eigenvalue of the resulting Jacobian. In the case of infinitely extended translation-invariant systems, the spatial dependence corresponds to harmonics $e^{\mathrm{i} kx}$ with wavenumber $k$. Introducing $\vecg u_0+ \chi\vecg{\hat u} e^{\lambda t + \mathrm{i} kx}$ with $\chi\ll1$ into the kinetic equation and linearizing in $\chi$ ultimately gives the dispersion relation $\lambda(k)$ and corresponding eigenmodes $\vecg{\hat u}$. The real part $\Delta:=\mathrm{Re}\lambda$ is a rate that characterizes the exponential growth ($\Delta>0$) or decay ($\Delta<0$) of the corresponding linear mode. The imaginary part $\Omega:=\mathrm{Im}\lambda$ corresponds to a frequency that can be zero (stationary case) or nonzero (oscillatory case).

\begin{table}[hbt]
\begin{tabular}{c || c | c}
& nonconserved dynamics& conserved dynamics\\
\hline
\hline
homogeneous/large-scale, stationary & Allen-Cahn (III$_\mathrm{s}$)& Cahn-Hilliard~(II$_\mathrm{s}$)\\
\hline
homogeneous/large-scale, oscillatory & Hopf\footnote{Also known as ``Poincar\'e-Andronov-Hopf''.} (III$_\mathrm{o}$)& conserved-Hopf~(II$_\mathrm{o}$)\\
\hline
small-scale, stationary & Turing (I$_\mathrm{s}$) & conserved-Turing (-)\\
\hline
small-scale, oscillatory & wave\footnote{Also called ``finite-wavelength Hopf'' or ``oscillatory Turing''.}  (I$_\mathrm{o}$)& conserved-wave (-)\\   
\hline
\end{tabular}
\caption{Naming convention of linear instabilities (and corresponding bifurcations) classified via their spatial (homogeneous/large-scale vs.\ small-scale) and temporal (stationary vs.\ oscillatory) properties for the cases of nonconserved and conserved dynamics. In parentheses we give the names in the (incomplete) classification of Cross and Hohenberg \cite{CrHo1993rmp}.}
\label{tab:instab}
\end{table}

There are eight basic types of instability when basing the classification on the combination of three features: (i) large-scale vs.\ small-scale instability, (ii) stationary vs.\ oscillatory instability, and (iii) instability without and with conservation law(s). Each of them features typical dominant modes directly at and in the vicinity of the instability threshold. Denoting the control parameter by $\varepsilon$, the left hand side panels of Fig.~\ref{fig:linstab} present the main types in our amended classification by showing the dispersion relation $\Delta(k)$. In each case we give relations below ($\varepsilon<0$), at ($\varepsilon=0$) and above ($\varepsilon>0$) the instability threshold. We also indicate the critical wavenumber $k_\mathrm{c}$ where at instability onset a maximum of $\Delta(k)$ touches zero, the marginal wavenumber(s) $k_\pm$ where $\mathrm{Re}\lambda(k)$ crosses zero above onset, and the fastest growing wavenumber $k_\mathrm{m}$ where $\Delta(k)$ has a maximum above onset. Note that for each of the four shown cases there exist a stationary and an oscillatory variant.

The right hand side panels of Fig.~\ref{fig:linstab} give the loci of $k_\pm$ and $k_\mathrm{m}$ in the ($k, \varepsilon$)-plane thereby illustrating the band of unstable wavenumbers in its dependence on $\varepsilon$. Note that in each case these dependencies are based on the leading order dispersion relation. When higher orders in $k^2$ are included $k_\pm(\varepsilon)$ and $k_\mathrm{m}(\varepsilon)$ can be quantitatively different for $\varepsilon\neq 0$. Our naming convention for the eight instabilities is given in Table~\ref{tab:instab}.
For reference, the classification of Cross and Hohenberg \cite{CrHo1993rmp} is given, but note that it only distinguishes six cases. Ref.~\cite{CrHo1993rmp} further states that ``type~II can often be scaled to resemble type~I''. In our opinion this is not correct. Also, their statements that on the one hand stationary and oscillatory instabilities have at onset frequencies $\Omega=0$ and $\Omega=\mathcal{O}(1)$, respectively, and on the other hand that type~II occurs in the presence of a conservation law seem to contradict each other, as the oscillatory case~II has $\Omega\sim\mathcal{O}(\varepsilon^2)$. In contrast, we propose a classification that takes the importance of conservation laws directly and systematically into account. Note, however, that in the cases with conservation law we restrict our attention to situations where the conservation law is directly related to the mode that becomes unstable. Cases where fields with conserved and fields with nonconserved or mixed dynamics interact in such a way that a mode representing nonconserved dynamics becomes unstable in the presence of an additional conservation law are not considered here. They could be part of a classification that incorporates the interaction of different branches of the dispersion relation.

Inspecting the first and third row of Fig.~\ref{fig:linstab} we see that there are four basic instability types for nonconserved systems: If the critical wavenumber, i.e., the marginal wavenumber at onset is zero ($k_\mathrm{c}=0$) the marginal mode is homogeneous (sometimes called global) as it synchronously affects each point of the domain without any spatial modulation. We refer to it as a homogeneous, uniform or global instability. If the corresponding critical frequency $\Omega_\mathrm{c}$ is zero [nonzero] the instability is stationary [oscillatory]. The corresponding amplitude equations for the stationary (Allen-Cahn instability) and the oscillatory  (Hopf instability) case are the Allen-Cahn equation
\begin{equation}
  \partial_t B =  \mathrm{sgn}(\varepsilon) B +\partial_{xx} B + \alpha B^2 - B^3
  \label{eq:ae:ac}
\end{equation}
and the complex Ginzburg-Landau equation
\begin{equation}
\partial_t A = \mathrm{sgn}(\varepsilon) A  + (1+ \mathrm{i} \kappa_i)  \partial_{xx} A- (1+ \mathrm{i} \alpha_i) |A|^2A
\label{eq:amp-lso}
\end{equation}
 respectively. \\
  Here, $B(x,t)$ is the spatially slowly varying real amplitude of the spatially homogeneous stationary mode while $A(x,t)$ is the spatially slowly varying complex amplitude of the spatially homogeneous oscillatory mode at $k=0$.
Note that a derivation of the universal amplitude equation results in specific coefficients for each term on the right hand side that depend on the corresponding original model. Furthermore, the space and time variables describe spatially and temporally slow dynamics, hence, amplitude equations are often referred to as envelope equations. Introducing a transformation of space, time and amplitude one can always eliminate three coefficients. Here and for the following cases we give these simplified universal equations in the supercritical cases. In the corresponding subcritical cases the cubic nonlinearity acts destabilizing, e.g., it occurs with a positive sign in Eq.~\eqref{eq:amp-lso}. Then, higher order stabilizing terms, e.g., a quintic nonlinearity, must be included to obtain a well behaved system. The $\mathrm{sgn}(\varepsilon)$ function occurs since it determines whether the amplitude linearly grows or decays.

In contrast, a nonzero marginal wavenumber at onset ($k_\mathrm{c}\neq0$) indicates a small-scale instability, i.e., an instability of finite wavelength. 
The amplitude equation in the stationary case (Turing instability) is the real Ginzburg-Landau equation
\begin{equation}
\partial_t A = \mathrm{sgn}(\varepsilon) A  + \partial_{xx} A - |A|^2A
\end{equation}
where $A(x,t)$ is the slowly varying complex amplitude of the spatially periodic stationary mode at $k=k_\mathrm{c}$. In the oscillatory case (wave instability) the coupled complex Ginzburg-Landau equations
\begin{align}
\partial_t A_1 &= \mathrm{sgn}(\varepsilon) A_1 - c \partial_{x} A_1 + (1+ \mathrm{i}\kappa_i)  \partial_{xx} A_1 - (1+ \mathrm{i}\alpha_i) |A_1|^2A_1 - (\beta_r+ \mathrm{i} \beta_i) |A_2|^2A_1 \nonumber\\
  \partial_t A_2 & = \mathrm{sgn}(\varepsilon) A_2 + c \partial_{x} A_2+  (1+ \mathrm{i} \kappa_i) \partial_{xx} A_2 - (1+ \mathrm{i}\alpha_i) |A_2|^2A_2 - (\beta_r+ \mathrm{i} \beta_i) |A_1|^2A_2
                 \label{eq:amp-ccGL}
\end{align}
emerge as amplitude equation where complex amplitudes $A_1$ and $A_2$ correspond to left and right traveling waves, respectively. These equations are only valid for small group velocity $c$ (see sec.~VI.E of \cite{ArKr2002rmp},  Section 7.1 of \cite{Hoyle2006}, Sections~1 \& 2 of \cite{WiMC2005n}). If this condition is not fulfilled a nonlocal equation is derived \cite{Knob1992}.
The four described cases for systems without conservation laws are all well covered by the Cross-Hohenberg classification (as types III$_\mathrm{s}$, III$_\mathrm{o}$, I$_\mathrm{s}$, I$_\mathrm{o}$, respectively) \cite{CrHo1993rmp} and are widely analyzed in the literature.

However, for systems with a conservation law there are another four basic types again distinguished based on wavenumber and mode type at onset. They are shown in the second and fourth row of Fig.~\ref{fig:linstab}. The conservation law results in the existence of a neutral mode at zero wavenumber, i.e., $\lambda=0$ at $k=0$ at all values of $\varepsilon$. Note that the instability thresholds are equivalent to the corresponding cases without conservation law. However, the fastest growing mode above onset behaves differently. For instance, in the case of zero marginal wavenumber at onset ($k_\mathrm{c}=0$, second row of Fig.~\ref{fig:linstab}, Cahn-Hilliard instability), $k_\mathrm{m}$ increases with $\varepsilon$ in contrast to the case of an Allen-Cahn instability where $k_\mathrm{m}=k_\mathrm{c}=0$. This implies that the instability is observed as a large-scale instability, not a homogeneous one. The growth of a homogeneous mode is incompatible with a fully conserved dynamics.
The corresponding amplitude equation has only recently been systematically derived \cite{BeRZ2018pre}. It corresponds to the Cahn-Hilliard equation (well known from other contexts)
\begin{equation}
  \partial_t B =  \partial_{xx} \left(-\mathrm{sgn}(\varepsilon) B + \alpha B^2+ B^3 - \partial_{xx} B\right)
  \label{eq:CHampl}
\end{equation}
where $B(x,t)$ is the spatially slowly varying real amplitude of the spatially homogeneous stationary mode at $k=0$. 
Since the Cahn-Hilliard equation is a continuity equation the total growth, i.e., the growth integrated over the domain, $\int {\mathrm d}x \partial_t B$, vanishes. This confirms that conserved quantities can only be spatially redistributed within the domain but overall do not change. Depending on the value of $\alpha$ (and the mean value of $B$ if not fixed to zero by an affine transformation) the resulting bifurcation can be sub- or supercritical.\footnote{The large-scale case is mentioned as type~II but not further discussed in the Cross-Hohenberg classification \cite{CrHo1993rmp} as ``type~II can often be scaled to resemble type~I''. However, the different amplitude equations show that this is actually not the case. This reflects that the existence of the conservation law changes central features of the system.} 

The small-scale cases with conservation law [Fig.~\ref{fig:linstab}~(g,h)] are not explicitly mentioned in the Cross-Hohenberg classification \cite{CrHo1993rmp}, but their relevance and distinct features became later more widely known \cite{MaCo2000n,WiMC2005n}. Here, we call the stationary and oscillatory case conserved-Turing and conserved-wave instability, respectively. In the stationary case the corresponding amplitude equation corresponds to a real Ginzburg-Landau equation coupled to a nonlinear diffusion equation (see \cite{MaCo2000n} and Section~9.4 of \cite{Hoyle2006})
\begin{align}
\partial_t A & = \mathrm{sgn}(\varepsilon) A  + \partial_{xx} A - |A|^2A - A B\nonumber\\
\partial_t B & = \partial_{xx}\left( \nu B + \mu  |A|^2\right)\,.
\end{align}
Here, $A$ is the slowly varying complex amplitude of a spatially periodic stationary mode and $B$ is the slowly varying real amplitude of the spatially homogeneous stationary mode. In the oscillatory case one finds two complex Ginzburg-Landau equations (for complex amplitudes $A_1$ and $A_2$ of left and right traveling waves, respectively) coupled to an equation for a real scalar mode of amplitude $B$ (see pg.~1034 of \cite{WiMC2005n}), i.e., 
\begin{align}
\partial_t A_1 &=\mathrm{sgn}(\varepsilon)A_1 - c \partial_{x} A_1 + (1+\mathrm{i} \kappa_i)  \partial_{xx} A_1 - (1+\mathrm{i}\alpha_i) |A_1|^2A_1 - (\beta_r+\mathrm{i} \beta_i) |A_2|^2A_1 - (1+\mathrm{i} \gamma_i) A_1B\nonumber\\
\partial_t A_2 & = \mathrm{sgn}(\varepsilon) A_2 + c \partial_{x} A_2 +  (1+\mathrm{i}\kappa_i)  \partial_{xx} A_2 - (1+\mathrm{i}\alpha_i) |A_2|^2A_2 - (\beta_r+\mathrm{i} \beta_i) |A_1|^2A_2 - (1+\mathrm{i} \gamma_i) A_2 B\nonumber\\
  \partial_t B & =  \partial_{xx}\left( \nu B + \mu  (|A_1|^2-|A_2|^2) \right).
        \label{eq:ae:gloc}
\end{align}
Again, these are only valid for small group velocity $c$.

Although the presented picture might at first sight seem complete, the careful reader will have noticed that Eqs.~\eqref{eq:ae:ac} to \eqref{eq:ae:gloc} only present the amplitude equations for seven of the eight cases of linear instabilities discussed above: we have neglected the large-scale oscillatory instability with conservation law (conserved-Hopf instability). To our knowledge, this case has not yet been systematically treated in the literature, and no corresponding amplitude equation has been derived yet. This is the subject of the main text with all details given in the next section of the Supplementary Material.

\subsection{Derivation of amplitude equation for conserved-Hopf instability}
\label{sec:app:consHopf}
To provide the detailed derivation of the general nonreciprocal Cahn-Hilliard model as an amplitude equation for the conserved-Hopf instability, i.e., a large-scale oscillatory instability in systems with two conservation laws, we consider a homogeneous reflection-symmetric system whose evolution is modeled by coupled kinetic equations for two conserved ($\rho$ and $\sigma$) and $N$ nonconserved ($\vecg{n}=(n_1,\dots,n_N)$) scalar field variables, i.e.,
\begin{align}\label{eq:2Conssystem}
\begin{split}
\partial_t \rho =& -\vec \nabla\cdot\left( Q(\vecg u) \vec \nabla \eta(\vecg u, \vec \nabla) \right)~\\
\partial_t \sigma =& -\vec \nabla\cdot \left( R(\vecg u) \vec \nabla \mu(\vecg u, \vec \nabla) \right)~\\
\partial_t \vecg{n} =& \vecg{F}(\vecg u, \vec \nabla),
\end{split}
\end{align}
From here onwards $\vecg u =\left(\rho, \sigma, \vecg{n}\right)$ is used as abbreviation where convenient.
The dynamics of the conserved quantities is determined by the divergence of the corresponding fluxes. Each flux is the product of a mobility function ($Q$ and $R$, respectively) and the gradient of a potential ($\eta$ and $\mu$, respectively). In general, these are nonequilibrium (chemical) potentials that can not be obtained as variational derivatives from a single underlying energy functional. This renders the system active. For scalar fields, reflection symmetry implies that Eqs.~\eqref{eq:2Conssystem} are invariant under the transformation $\vec x\to -\vec x$. Therefore, the individual terms within the potentials $\eta$ and $\mu$ do either not include derivatives or an even number of them. Normally, the mobilities are assumed to be functions of the fields, but terms with an even number of derivatives may also be easily accommodated. Furthermore, at least one uniform steady state $\vecg u_0$ with $\vecg{F}(\vecg u_0)=0$ shall exist.
The potentially simplest example for a system~\eqref{eq:2Conssystem} is a reaction diffusion system with $N+2$ species that has been rearranged to explicitly show the two conservation laws. An example with $N=1$ is considered in Section~\ref{sec:app:rd}. However, the structure~\eqref{eq:2Conssystem} also captures much more complex systems, e.g., all examples mentioned in the main text.

We consider the case of a conserved-Hopf instability, i.e., at control parameter $\delta=0$ the considered $\vecg u_0$ state becomes unstable with a dispersion relation above onset as in Eq.~(1) and Fig.~1 of the main text. Note that in a system with $N\ge2$ nonconserved quantities a standard Hopf instability is also still possible. However, then the conserved quantities do not contribute to the corresponding linear modes, and interactions between nonconserved oscillatory and conserved real modes will only occur nonlinearly. Here, such a setting is not considered. The conserved-Hopf instability we are here interested in always involves the branches of the dispersion relation containing the two neutral modes at $k=0$.

      \begin{table}[htb]
	\begin{tabular}{| c | c |}
		\hline
		Quantities in& ~\\
		\hline
		Eq.~\eqref{eq:quant_expand} & 
		{$\!\begin{aligned} 
			Q_0 &= Q(\vecg u_0) \\    
			Q_1 &= \sum_i \frac{\partial Q}{\partial u_i}\bigg|_{\vecg u_0}\left(\vecg u_1 \right)_i ~\\
			Q_2 &= \frac12\sum_{i,j} \frac{\partial^2 Q}{\partial u_i \partial u_j}\bigg|_{\vecg u_0}\left(\vecg u_1 \right)_i\left(\vecg u_1 \right)_j + \sum_i \frac{\partial Q}{\partial u_i}\bigg|_{\vecg u_0}\left(\vecg u_2 \right)_i~\\
			&R_m, \, \eta_m, \, \mu_m,\, \vecg F_m \,\,\, \text{for $m=0,1,2$ analogously}~\\
			\eta_3 &=  \frac16 \sum_{i,j,k}  \frac{\partial^3 \eta}{\partial u_i \partial u_j \partial u_k}\bigg|_{\vecg u_0}\left(\vecg u_1 \right)_i\left(\vecg u_1 \right)_j\left(\vecg u_1 \right)_k +   \sum_i \frac{\partial \eta}{\partial \left( \vec \nabla^2 u_i\right)}\bigg|_{\vecg u_0}\left(\vec \nabla^2 \vecg u_1 \right)_i~\\& +  \sum_{i,j} \frac{\partial^2 \eta}{\partial u_i \partial u_j}\bigg|_{\vecg u_0}\left(\vecg u_1 \right)_i\left(\vecg u_2 \right)_j + \sum_i \frac{\partial \eta}{\partial u_i}\bigg|_{\vecg u_0}\left(\vecg u_3 \right)_i~\\
			&\mu_3,\, \vecg F_3 \,\,\, \text{ analogously}~\\
			\end{aligned}$}
		\\
		\hline
		Eq.~\eqref{eq:eta1} & 
		{$\!\begin{aligned}   
			\eta_\rho &= \frac{\partial \eta}{\partial \rho}\bigg|_{\vecg u_0} + \sum_i \frac{\partial \eta}{\partial n_i}\bigg|_{\vecg u_0} \left(\vecg{n_\rho}\right)_i ~\\
			\eta_\sigma &= \frac{\partial \eta}{\partial \sigma}\bigg|_{\vecg u_0} + \sum_i \frac{\partial \eta}{\partial n_i}\bigg|_{\vecg u_0} \left(\vecg{n_\sigma}\right)_i ~\\
			&\mu_\rho,\, \mu_\sigma \,\,\, \text{ analogously}~\\
			\end{aligned}$}	\\
		\hline
		Eq.~\eqref{eq:QRetamu} & 		
		{$\!\begin{aligned}   
			q_\rho &= \frac{\partial Q}{\partial \rho}\bigg|_{\vecg u_0} + \sum_i \frac{\partial Q}{\partial n_i}\bigg|_{\vecg u_0} \left(\vecg{n_\rho}\right)_i ~\\
			q_\sigma &= \frac{\partial Q}{\partial \sigma}\bigg|_{\vecg u_0} + \sum_i \frac{\partial Q}{\partial n_i}\bigg|_{\vecg u_0} \left(\vecg{n_\sigma}\right)_i ~\\
			&r_\rho,\, r_\sigma \,\,\, \text{ analogously}~\\
						\eta_{\rho \rho}= & \frac12 \frac{\partial^2 \eta}{\partial \rho^2}\bigg|_{\vecg u_0} + \sum_i \frac{\partial^2 \eta}{\partial \rho \partial n_i}\bigg|_{\vecg u_0} \left(\vecg{n_\rho}\right)_i + \frac12 \sum_{i,j} \frac{\partial^2 \eta}{\partial n_i \partial n_j}\bigg|_{\vecg u_0} \left(\vecg{n_\rho}\right)_i \left(\vecg{n_\rho}\right)_j + \textcolor{red}{\sum_i \frac{\partial \eta}{\partial n_i}\bigg|_{\vecg u_0} \left(\vecg{n_{\rho \rho}}\right)_i} ~\\
						~
			\eta_{\rho \sigma}= &  \frac{\partial^2 \eta}{\partial \rho \partial \sigma}\bigg|_{\vecg u_0} + \sum_i \frac{\partial^2 \eta}{\partial \rho \partial n_i}\bigg|_{\vecg u_0} \left(\vecg{n_\sigma}\right)_i + \sum_i \frac{\partial^2 \eta}{\partial \sigma \partial n_i}\bigg|_{\vecg u_0} \left(\vecg{n_\rho}\right)_i + \sum_{i,j} \frac{\partial^2 \eta}{\partial n_i \partial n_j}\bigg|_{\vecg u_0} \left(\vecg{n_\rho}\right)_i \left(\vecg{n_\sigma}\right)_j~\textcolor{red}{+ \sum_i \frac{\partial \eta}{\partial n_i}\bigg|_{\vecg u_0} \left(\vecg{n_{\rho \sigma}}\right)_i}\\		
~			
				\eta_{\sigma \sigma}= & \frac12 \frac{\partial^2 \eta}{\partial \sigma^2}\bigg|_{\vecg u_0} + \sum_i \frac{\partial^2 \eta}{\partial \sigma \partial n_i}\bigg|_{\vecg u_0} \left(\vecg{n_\sigma}\right)_i + \frac12 \sum_{i,j} \frac{\partial^2 \eta}{\partial n_i \partial n_j}\bigg|_{\vecg u_0} \left(\vecg{n_\sigma}\right)_i \left(\vecg{n_\sigma}\right)_j~\textcolor{red}{+ \sum_i \frac{\partial \eta}{\partial n_i}\bigg|_{\vecg u_0} \left(\vecg{n_{\sigma \sigma}}\right)_i}\\
			&\mu_{\rho \rho},\, \mu_{\rho \sigma}, \, \mu_{\sigma \sigma} \,\,\, \text{ analogously}~\\
			\end{aligned}$}	\\
		\hline
				Eq.~\eqref{eq:order5_2} & 		
		{$\!\begin{aligned}   
q_{\rho \rho}= & \frac12 \frac{\partial^2 Q}{\partial \rho^2}\bigg|_{\vecg u_0} + \sum_i \frac{\partial^2 Q}{\partial \rho \partial n_i}\bigg|_{\vecg u_0} \left(\vecg{n_\rho}\right)_i + \frac12 \sum_{i,j} \frac{\partial^2 Q}{\partial n_i \partial n_j}\bigg|_{\vecg u_0} \left(\vecg{n_\rho}\right)_i \left(\vecg{n_\rho}\right)_j + \sum_i \frac{\partial Q}{\partial n_i}\bigg|_{\vecg u_0} \left(\vecg{n_{\rho \rho}}\right)_i\\			
			&q_{\rho \sigma}, \, q_{\sigma \sigma}, \, r_{\rho \rho},\, r_{\rho \sigma}, \, r_{\sigma \sigma} \,\,\, \text{ analogously}~\\
			\eta_{\rho \rho \rho}= & \frac16 \frac{\partial^3 \eta}{\partial \rho^3}\bigg|_{\vecg u_0} + \frac12 \sum_i \frac{\partial^3 \eta}{\partial \rho^2 \partial n_i}\bigg|_{\vecg u_0} \left(\vecg{n_\rho}\right)_i + \frac12 \sum_{i,j} \frac{\partial^3 \eta}{\partial \rho \partial n_i \partial n_j}\bigg|_{\vecg u_0} \left(\vecg{n_\rho}\right)_i \left(\vecg{n_\rho}\right)_j + \frac16 \sum_{i,j,k} \frac{\partial^3 \eta}{ \partial n_i \partial n_j \partial n_k}\bigg|_{\vecg u_0} \left(\vecg{n_\rho}\right)_i \left(\vecg{n_\rho}\right)_j \left(\vecg{n_\rho}\right)_k~~~\\ &
			 + \sum_i \frac{\partial^2 \eta}{\partial \rho \partial n_i}\bigg|_{\vecg u_0} \left(\vecg{n_{\rho \rho}}\right)_i + \sum_{i,j} \frac{\partial^2 \eta}{\partial n_i \partial n_j}\bigg|_{\vecg u_0} \left(\vecg{n_{\rho \rho}}\right)_i \left(\vecg{n_\rho}\right)_j + \sum_i \frac{\partial \eta}{\partial n_i}\bigg|_{\vecg u_0} \left(\vecg{n_{\rho \rho \rho}}\right)_i \\
			\eta_{\rho \rho \sigma}= & \frac12 \frac{\partial^3 \eta}{\partial \rho^2 \partial \sigma}\bigg|_{\vecg u_0} + \frac12 \sum_i \frac{\partial^3 \eta}{\partial \rho^2 \partial n_i}\bigg|_{\vecg u_0} \left(\vecg{n_\sigma}\right)_i + \sum_i \frac{\partial^3 \eta}{\partial \rho \partial \sigma \partial n_i}\bigg|_{\vecg u_0} \left(\vecg{n_\rho}\right)_i
			 + \sum_{i,j} \frac{\partial^3 \eta}{\partial \rho \partial n_i \partial n_j}\bigg|_{\vecg u_0} \left(\vecg{n_\rho}\right)_i \left(\vecg{n_\sigma}\right)_j ~\\
			 & + \frac12 \sum_{i,j} \frac{\partial^3 \eta}{\partial \sigma \partial n_i \partial n_j}\bigg|_{\vecg u_0} \left(\vecg{n_\rho}\right)_i \left(\vecg{n_\rho}\right)_j + \frac12 \sum_{i,j,k} \frac{\partial^3 \eta}{ \partial n_i \partial n_j \partial n_k}\bigg|_{\vecg u_0} \left(\vecg{n_\rho}\right)_i \left(\vecg{n_\rho}\right)_j \left(\vecg{n_\sigma}\right)_k + \sum_i \frac{\partial^2 \eta}{\partial \rho \partial n_i}\bigg|_{\vecg u_0} \left(\vecg{n_{\rho \sigma}}\right)_i~\\
&  + \sum_i \frac{\partial^2 \eta}{\partial \sigma \partial n_i}\bigg|_{\vecg u_0} \left(\vecg{n_{\rho \rho}}\right)_i + \sum_{i,j} \frac{\partial^2 \eta}{\partial n_i \partial n_j}\bigg|_{\vecg u_0} \left(\vecg{n_{\rho \sigma}}\right)_i \left(\vecg{n_\sigma}\right)_j~ + \sum_i \frac{\partial \eta}{\partial n_i}\bigg|_{\vecg u_0} \left(\vecg{n_{\rho \rho \sigma}}\right)_i			 
			 \\			
			&\eta_{\rho \sigma \sigma},\, \eta_{\sigma \sigma \sigma}, \,\mu_{\rho \rho \rho},\, \mu_{\rho \rho \sigma}, \, \mu_{\rho \sigma \sigma}, \, \mu_{\sigma \sigma \sigma} \,\,\, \text{ analogously}~\\
			~
			\eta_{\rho x x} =&\frac{\partial \eta}{\partial \left( \vec \nabla^2 \rho\right)}\bigg|_{\vecg u_0} + \sum_i \frac{\partial \eta}{\partial \left( \vec \nabla^2 n_i\right)}\bigg|_{\vecg u_0}\left(\vecg n_\rho \right)_i  + \sum_i \frac{\partial \eta}{\partial n_i}\bigg|_{\vecg u_0} \left(\vecg{n_{\rho xx}}\right)_i~\\
				~
			\mu_{\sigma x x} =&\frac{\partial \mu}{\partial \left( \vec \nabla^2 \sigma\right)}\bigg|_{\vecg u_0} + \sum_i \frac{\partial \mu}{\partial \left( \vec \nabla^2 n_i\right)}\bigg|_{\vecg u_0}\left(\vecg n_\sigma \right)_i  + \sum_i \frac{\partial \mu}{\partial n_i}\bigg|_{\vecg u_0} \left(\vecg{n_{\sigma xx}}\right)_i~\\
			&\eta_{\sigma x x}, \,\mu_{\rho x x} \,\,\, \text{analogously}~\\
			\end{aligned}$}	\\
			\hline
	\end{tabular}
	\caption{Relations between the quantities in the original equation~\eqref{eq:2Conssystem} and the various coefficients appearing in equations \eqref{eq:quant_expand}, \eqref{eq:eta1}, \eqref{eq:QRetamu}, and \eqref{eq:order5_2} at the different orders in $\varepsilon$.}
	\label{tab:compare_coeff}
\end{table}

To perform the weakly nonlinear analysis valid in the vicinity of instability onset, i.e., for $\delta=\delta_2 \varepsilon^2$, we expand all fields in the smallness parameter $\varepsilon\ll 1$, i.e.,
\begin{align}\label{eq:fields_expand}
\vecg u  (\vec X,\tau,T)= \vecg u_0 + \varepsilon \vecg u_1(\vec X,\tau,T) + \varepsilon^2 \vecg u_2(\vec X,\tau,T) + \dots,
\end{align}
take account of the scaling discussed in the main text by writing $\vec \nabla_{\!\vec x} = \varepsilon \vec \nabla_{\!\vec X}$ and $\partial_t= \varepsilon^2 \partial_\tau + \varepsilon^4 \partial_T$, and also expand all quantities occurring on the right hand side of Eqs.~\eqref{eq:2Conssystem} as
\begin{align}\label{eq:quant_expand}
\begin{split}
Q=& Q_0 + \varepsilon Q_1 + \varepsilon^2 Q_2 + \dots~\\
R=& R_0 + \varepsilon R_1 + \varepsilon^2 R_2 + \dots~\\
\eta=& \eta_0 + \varepsilon \eta_1 + \varepsilon^2 \eta_2 +  \varepsilon^3 \eta_3 +\dots~\\
\mu=& \mu_0 + \varepsilon \mu_1 + \varepsilon^2 \mu_2 +  \varepsilon^3 \mu_3 + \dots~~\\
\vecg F =& \varepsilon \vecg F_1 + \varepsilon^2 \vecg F_2+\varepsilon^3 \vecg F_3+ \dots\, ,
\end{split}
\end{align}
where $\vecg F_0 = \vecg F(\vecg u_0)=0$ was used. The various coefficients are given by corresponding Taylor expansions, e.g., $Q_2= \vecg u_1 \cdot \frac12\frac{\partial^2 Q}{\partial \vecg u^2}\big|_{\vecg u_0} \cdot \vecg u_1  + \frac{\partial Q}{\partial \vecg u}\big|_{\vecg u_0} \cdot \vecg u_2$. Examples for these coefficients are given in Table~\ref{tab:compare_coeff}.
All expansions and scalings are inserted into the model equations \eqref{eq:2Conssystem}, that are then considered order by order in $\varepsilon$. Due to the scaling implied by the dispersion relation, one needs to successively consider all orders up to $\mathcal{O}(\varepsilon^5)$ to obtain evolution equations that capture dynamics effects on the slow timescale $T$.

The general procedure to follow at each order $i=1,\dots,5$ is: First, determine the nonconserved fields $\vecg n_i$ as (nonlinear) functions of the conserved fields $\sigma$ and $\rho$, hence, consider the dynamics of the nonconserved fields that is slaved to the dynamics of the conserved fields. Second, obtain the continuity equations to the corresponding order by inserting the expressions for $\vecg n_i$ into the appropriate mobilities $Q_i$, $R_i$ and potentials $\eta_i$, $\mu_i$. Finally, the dynamics for $\rho$ and $\sigma$ are combined into two coupled equations including terms up to $\partial_T \rho_1$ and $\partial_T \sigma_1$, respectively, that represent the sought-after amplitude equation.
Now we proceed order by order.

\paragraph{Order $\varepsilon$:} 
As expected, we recover the linear result at $k=0$: The contributions of the two continuity equations vanish and the remaining $N$ equations become the algebraic system
\begin{align}
  0= \vecg F_1 \label{eq:order1}
 \end{align}
linear in $N+2$ unknown quantities $\vecg u_1$. It is solved for
\begin{align}\label{eq:n1}
\vecg n_1(\vec X,\tau,T) = \vecg n_\rho \rho_1(\vec X,\tau,T) + \vecg n_\sigma \sigma_1(\vec X,\tau,T)
\end{align}
where $\vecg n_\rho$ and $\vecg n_\sigma$ correspond to the zero eigenmodes $(1,0,\vecg n_\rho)$ and $(0,1,\vecg n_\sigma)$, respectively, discussed in the main text.

\paragraph{Order $\varepsilon^2$:} 
As in $\mathcal{O}(\varepsilon)$, the continuity equations are trivially fulfilled and the nonconserved dynamics
give
\begin{align}
  0=  \vecg F_2\,. \label{eq:order2}
\end{align}
These algebraic relations consist of a linear part similar to Eq.~\eqref{eq:n1} but for the fields $\vecg u_2$ and a nonlinear part
quadratic in $\rho_1$ and $\sigma_1$ (after eliminating quadratic parts involving $\vecg n_1$ via Eq.~\eqref{eq:n1}). The nonlinearities correspond to the inhomogeneity of the algebraic system for the $\vecg u_2$.
In consequence, $\vecg n_2$ is now given as a sum of a part linear in the amplitudes $\rho_2$ and $\eta_2$ and the nonlinearity, namely,
\begin{equation}\label{eq:n2}
\vecg n_2 = \vecg n_\rho \rho_2+ \vecg n_\sigma \sigma_2 + \vecg n_{\rho\rho} \rho_1^2 + \vecg n_{\rho\sigma} \rho_1 \sigma_1 +\vecg n_{\sigma \sigma} \sigma_1^2\,.
\end{equation} 
Here and in the following, the vectors of real constant coefficients $\vecg n_{\alpha}$, $\vecg n_{\alpha\beta}$, etc.\ depend on the specific functions $\vecg F$. 

\paragraph{Order $\varepsilon^3$:} 
At the next order, the first nonvanishing contribution from the continuity equations appears, resulting in the system
\begin{align}\label{eq:order3}
 \begin{split}
&\partial_\tau \rho_1 = - \vec \nabla_{\!\vec{X}} \cdot \left(Q_0 \vec \nabla_{\!\vec{X}}\, \eta_1\right)~\\
& \partial_\tau \sigma_1 = - \vec \nabla_{\!\vec{X}}\cdot \left( R_0 \vec \nabla_{\!\vec{X}}\, \mu_1\right)~\\
& \partial_\tau \vecg n_1= \vecg F_3\,,
 \end{split}
\end{align}
where $\eta_1 $ and $\mu_1$ are linear in $\rho_1$, $\sigma_1$ and $\vecg n_1$. For the latter we insert Eq.~\eqref{eq:n1} and obtain
\begin{align}~\label{eq:eta1}
	\begin{split}
\eta_1 &= \eta_\rho \rho_1 + \eta_\sigma \sigma_1\\	
\mu_1 &= \mu_\rho \rho_1 + \mu_\sigma \sigma_1  		
\end{split}
\end{align}
with real constant coefficients $\eta_\rho$, $\eta_\sigma$, $\mu_\rho$ and $\mu_\sigma$ as exemplified in Table~\ref{tab:compare_coeff} that also provides further coefficients appearing at higher orders in $\varepsilon$.
Further, $Q_0$, $R_0$ are constants and inserting the expressions~\eqref{eq:eta1} we can write the first two equations in~\eqref{eq:order3} as a linear system
\begin{align}\label{eq:leading_osc0}
\partial_\tau \left(\begin{array}{c}
\rho_1\\
\sigma_1
\end{array}\right)
= -\left(\begin{array}{c c}
 Q_0 \eta_\rho & Q_0 \eta_\sigma~\\
R_0 \mu_\rho & R_0 \mu_\sigma
\end{array}\right)
\vec \nabla_{\!\vec X}^2 \left(\begin{array}{c}
\rho_1\\
\sigma_1
\end{array}\right)\,.
\end{align}
Applying $(\rho_1, \sigma_1) \sim \Exp{\mathrm{i} \vec k  \cdot\vec X + \lambda \tau}$, its eigenvalues are
\begin{equation}\label{eq:lambda_leading_Hopf}
\lambda_\pm =k^2   \frac{Q_0 \eta_\rho + R_0 \mu_\sigma}{2}   \pm  k^2 \sqrt{\frac{(Q_0 \eta_\rho -R_0 \mu_\sigma)^2}{4} + Q_0 \eta_\sigma R_0 \mu_\rho  }\, ,
\end{equation}
where $k=|\vec k|$.
Comparing to the dispersion relation (Eq.~(1) of the main text) allows us to identify
\begin{align}\label{eq:delta_omega}
\begin{split}
\delta =& \frac{Q_0 \eta_\rho + R_0 \mu_\sigma}{2}~\\
\mathrm{i} \omega =& \sqrt{\frac{(Q_0 \eta_\rho -R_0 \mu_\sigma)^2}{4} + Q_0 \eta_\sigma R_0 \mu_\rho  }\,.
\end{split}
\end{align}
At onset of the conserved-Hopf instability the growth rate $\delta$ vanishes, i.e.,

  \begin{equation}\label{eq:onset_cond}
Q_0\eta_\rho =- R_0 \mu_\sigma
\end{equation} 
and the eigenvalues are purely imaginary, i.e.,
\begin{equation}\label{eq:inequality}
\frac{(Q_0 \eta_\rho -R_0 \mu_\sigma)^2}{4} + Q_0 \eta_\sigma R_0 \mu_\rho <0\,.
\end{equation}
Using~\eqref{eq:onset_cond}, this implies
\begin{equation}
\eta_\sigma  \mu_\rho< \eta_\rho \mu_\sigma <0,
\end{equation}
and thereby defines a nonreciprocity condition for the linear coupling terms within the potentials. Since we are interested in the dynamics closely above [below] the instability onset where $\delta=\delta_2 \varepsilon^2$ with $\delta_2>0$ [$\delta_2<0$], Eq.~\eqref{eq:onset_cond} only holds at leading order, i.e., including the next order we have $Q_0\eta_\rho= - R_0 \mu_\sigma + 2 \delta_2\varepsilon^2$.
This makes the result fully consistent with the scaling based on Eqs.~(1) of the main text, as the oscillations of leading order frequency $\omega k^2$ occur on the timescale $\tau$ [given that $\omega=\mathcal{O}(1)$] where the growth rate vanishes. Growth only occurs on the slower time scale $T$. Specifically, the $\mathcal{O}(\varepsilon^2)$ contribution in $\delta$ is then considered when below  discussing order $\varepsilon^5$ terms. Here, at $\mathcal{O}(\varepsilon^3)$, we can simply set $\delta=0$. With this, we now reformulate Eq.~\eqref{eq:leading_osc0}.
Note that any linear combination of $\rho$ and $\sigma$, e.g.,
\begin{align}\label{eq:AB}
A(\vec X,\tau,T) = a_\rho \rho(\vec X,\tau,T) + a_\sigma \sigma(\vec X,\tau,T)~\\
B(\vec X,\tau,T) = b_\rho \rho(\vec X,\tau,T) + b_\sigma \sigma(\vec X,\tau,T)\,.
\end{align}
with coefficients $a_\rho$, $a_\sigma$, $b_\rho$ and $b_\sigma$ 
results in an equivalent formulation for alternative conserved fields $A$ and $B$. We use the resulting freedom to simplify Eq.~\eqref{eq:leading_osc0}
by choosing
\begin{equation}\label{eq:AB_coeff}
b_\sigma = -\frac{ \omega a_\rho +  Q_0 \eta_\rho b_\rho }{R_0 \mu_\rho }\, , \quad a_\sigma = \frac{ \omega b_\rho -Q_0 \eta_\rho  a_\rho }{R_0 \mu_\rho}
\end{equation}
and $a_\rho$, $b_\rho$ are normalization constants.
This reduces Eqs.~\eqref{eq:leading_osc0} to
\begin{align}\label{eq:leading_oscAB}
\partial_\tau \left(\begin{array}{c}
A_1~\\
B_1
\end{array}\right)
= \left(\begin{array}{c c}
0 & -\omega~\\
\omega & 0
\end{array}\right)
\vec \nabla_{\!\vec X}^2 \left(\begin{array}{c}
A_1~\\
B_1
\end{array}\right),
\end{align}
i.e., the leading order oscillation is represented by an antisymmetric dynamic coupling of $A$ and $B$, that represents the lowest order nonreciprocal coupling. The form of Eq.~\eqref{eq:leading_oscAB} allows one to easily show that the corresponding part of the dynamics is not dissipative since it is equivalent to the structure of the Schr\"odinger equation for a free particle (similar to the complex Ginzburg-Landau equation without the local terms).

For simplicity of notation, here, we proceed with the description using the original quantities $\rho$ and $\sigma$. It helps us to identify the structure of the two continuity equations in terms of $Q$, $R$, $\eta$ and $\mu$. 
Still at $\mathcal{O}(\varepsilon^3)$, we determine $\vecg n_3$ via the third equation in \eqref{eq:order3}: we insert $\vecg n_1$ from Eq.~\eqref{eq:n1} and use Eqs.~\eqref{eq:leading_osc0} for the time derivative to obtain
\begin{align}
\begin{split}\label{eq:n3}
\vecg n_3 = & \vecg n_\rho \rho_3+ \vecg n_\sigma \sigma_3 + \vecg n_{\rho\rho} 2 \rho_1 \rho_2 + \vecg n_{\rho\sigma} \left(\rho_1 \sigma_2 + \rho_2 \sigma_1\right) +\vecg n_{\sigma \sigma} 2 \sigma_1 \sigma_2
~\\
& + \vecg n_{\rho\rho\rho} \rho_1^3  + \vecg n_{\rho\rho\sigma} \rho_1^2 \sigma_1 + \vecg n_{\rho\sigma\sigma} \rho_1 \sigma_1^2+ \vecg n_{\sigma \sigma \sigma} \sigma_1^3 ~\\
& + \vecg n_{\rho xx} \vec \nabla_{\!\vec X}^2 \rho_1 +  \vecg n_{\sigma xx} \vec \nabla_{\!\vec X}^2 \sigma_1\,.
\end{split}
\end{align}
Similar to the expression for $\vecg n_2$ it consists of the sum of a linear part given by Eq.~\eqref{eq:n1} applied to $\vecg u_3$ and a nonlinear part that corresponds to an inhomogeneity.

\paragraph{Order $\varepsilon^4$:} 
At the next order we obtain
\begin{align}\label{eq:order4}
 \begin{split}
& \partial_\tau \rho_2 = - \vec \nabla_{\!\vec X} \cdot \left(Q_0 \vec \nabla_{\!\vec X} \eta_2\right) - \vec \nabla_{\!\vec X} \cdot \left(Q_1 \vec \nabla_{\!\vec X} \eta_1\right)~\\
& \partial_\tau \sigma_2 = - \vec \nabla_{\!\vec X} \cdot \left(R_0 \vec \nabla_{\!\vec X} \mu_2\right) - \vec \nabla_{\!\vec X} \cdot \left(R_1 \vec \nabla_{\!\vec X} \mu_1\right)~\\
& \partial_\tau \vecg n_2= \vecg F_4
\end{split}
\end{align}
where additionally to the already known $Q_0$, $R_0$, $\eta_1$ and $\mu_1$ the higher order quantities $Q_1$, $R_1$, $\eta_2$ and $\mu_2$ enter. Using Eq.~\eqref{eq:n2} for $\vecg n_2$ we obtain
\begin{align}\label{eq:QRetamu}
\begin{split}
Q_1 = & q_\rho \rho_1 + q_\sigma \sigma_1 ~\\
R_1 = & r_\rho \rho_1 + r_\sigma \sigma_1 ~\\
\eta_2 = &\eta_\rho \rho_2 + \eta_\sigma \sigma_2  + \eta_{\rho \rho} \rho_1^2 + \eta_{\rho \sigma} \rho_1 \sigma_1 + \eta_{\sigma \sigma} \sigma_1^2   ~\\
\mu_2 = & \mu_\rho \rho_2 + \mu_\sigma \sigma_2 +  \mu_{\rho \rho} \rho_1^2 + \mu_{\rho \sigma} \rho_1 \sigma_1 + \mu_{\sigma \sigma} \sigma_1^2  
\end{split}
\end{align}

Furthermore, at order $\varepsilon^4$ one may also determine an algebraic expression for $\vecg n_4$. However, here, we do not present it because it does not contribute to the relevant leading order dynamics of $\rho$ and $\sigma$.

\paragraph{Order $\varepsilon^5$:} 
Finally, the fifth order gives the kinetic equations 
\begin{align}\label{eq:order5}
 \begin{split}
& \partial_T \rho_1 + \partial_\tau \rho_3 = - \vec \nabla_{\!\vec X} \cdot \left(Q_0 \vec \nabla_{\!\vec X} \eta_3\right) -  \vec \nabla_{\!\vec X}\cdot \left(Q_1 \vec \nabla_{\!\vec X} \eta_2 \right)- \vec \nabla_{\!\vec X}\cdot \left(Q_2 \vec \nabla_{\!\vec X} \eta_1\right) ~\\
& \partial_T \sigma_1 +\partial_\tau \sigma_3 = - \vec \nabla_{\!\vec X} \cdot \left(R_0 \vec \nabla_{\!\vec X} \mu_3\right) - \vec \nabla_{\!\vec X} \cdot \left(R_1 \vec \nabla_{\!\vec X} \mu_2\right)- \vec \nabla_{\!\vec X} \cdot \left(R_2 \vec \nabla_{\!\vec X} \mu_1\right)~\\
& \partial_T \vecg n_1 + \partial_\tau \vecg n_3= \vecg F_5\,.
 \end{split} 
\end{align}
Using the already determined expressions for $\vecg n_1$, $\vecg n_2$ and $\vecg n_3$ the complete right hand sides of the continuity Eqs.~\eqref{eq:order5} can be written as nonlinear functions of the $\rho_i$ and $\sigma_i$ with coefficients defined in a similar way as at lower orders:
\begin{align}\label{eq:order5_2}
\begin{split}
Q_2 = & q_\rho \rho_2 + q_\sigma \sigma_2 + q_{\rho \rho}  \rho_1^2 + q_{\rho \sigma} \rho_1 \sigma_1 + q_{\sigma \sigma}  \sigma_1^2 ~\\
R_2 = & r_\rho \rho_2 + r_\sigma \sigma_2 +r_{\rho \rho}  \rho_1^2 + r_{\rho \sigma} \rho_1 \sigma_1 + r_{\sigma \sigma}  \sigma_1^2 ~\\
\eta_3 = &\eta_\rho \rho_3 + \eta_\sigma \sigma_3  + 2 \eta_{\rho \rho}  \rho_1\rho_2 + \eta_{\rho \sigma} \left(\rho_1 \sigma_2 + \rho_2 \sigma_1\right) + 2\eta_{\sigma \sigma}  \sigma_1 \sigma_2   ~\\
& + \eta_{\rho \rho \rho} \rho_1^3 +  \eta_{\rho \rho \sigma} \rho_1^2 \sigma_1 +  \eta_{\rho \sigma \sigma} \rho_1 \sigma_1^2 + \eta_{\sigma \sigma \sigma} \sigma_1^3
 + \eta_{\rho xx} \vec \nabla_{\!\vec X}^2 \rho_1 +\eta_{\sigma xx} \vec \nabla_{\!\vec X}^2 \sigma_1~\\
~%
\mu_3 = & \mu_\rho \rho_3 + \mu_\sigma \sigma_3 + 2 \mu_{\rho \rho}  \rho_1\rho_2 + \mu_{\rho \sigma} \left(\rho_1 \sigma_2 + \rho_2 \sigma_1\right) + 2\mu_{\sigma \sigma} \sigma_1\sigma_2~ \\
& + \mu_{\rho \rho \rho} \rho_1^3 +  \mu_{\rho \rho \sigma} \rho_1^2 \sigma_1 +  \mu_{\rho \sigma \sigma} \rho_1 \sigma_1^2 + \mu_{\sigma \sigma \sigma} \sigma_1^3 
+  \mu_{\rho xx} \vec \nabla_{\!\vec X}^2 \rho_1 +\mu_{\sigma xx} \vec \nabla_{\!\vec X}^2 \sigma_1
\end{split}
\end{align}
This provides the weakly nonlinear expression for the time evolution
on the timescale $T$. To obtain the final amplitude equations we
combine the dynamics found at the different orders. In other words, we
recombine the different orders of the expansion of the fields into the
appropriate fields $\vecg u$ as the deviations from the steady state $\vecg u_0$. Specifically, we introduce $\varrho \equiv \rho - \rho_0$ and $\varsigma\equiv \sigma - \sigma_0$, i.e., the spatial and temporal modulations in $\rho(\vec x, t)$ and $\sigma(\vec x, t)$ away from their respective mean values $\rho_0$ and $\sigma_0$.

For instance, the dynamics of $\varrho$ is given by
\begin{align}\label{eq:combination}
\begin{split}
\partial_t \varrho = & \varepsilon^3 \partial_\tau \rho_1+ \varepsilon^4 \partial_\tau \rho_2 + \varepsilon^5 \left( \partial_T \rho_1 + \partial_\tau \rho_3 \right) + \mathcal{O}(\varepsilon^6) ~\\
=& -\varepsilon^3 \vec \nabla_{\!\vec X} \cdot \left(Q_0 \vec \nabla_{\!\vec X} \eta_1\right) - \varepsilon^4 \vec \nabla_{\!\vec X} \cdot \left(Q_0 \vec \nabla_{\!\vec X} \eta_2 + Q_1 \vec \nabla_{\!\vec X} \eta_1  \right) \\
& \, \, - \varepsilon^5 \vec \nabla_{\!\vec X} \cdot \left(Q_0 \vec \nabla_{\!\vec X} \eta_3 + Q_1 \vec \nabla_{\!\vec X} \eta_2  + Q_2 \vec \nabla_{\!\vec X} \eta_1\right) + \mathcal{O}(\varepsilon^6)~\\
=& - \varepsilon^2 \vec \nabla_{\!\vec X} \cdot \left( \left( Q_0 + \varepsilon Q_1 + \varepsilon^2 Q_2  \right) \vec \nabla_{\!\vec X} \left( \varepsilon \eta_1 + \varepsilon^2 \eta_2 + \varepsilon^3 \eta_3 \right) \right) + \mathcal{O}(\varepsilon^6).
\end{split}
\end{align}
Note that in the last step we have added selected terms, e.g.~$-\varepsilon^6 \vec \nabla_{\!\vec X} \cdot \left(Q_2 \vec \nabla_{\!\vec X} \eta_2\right)$ that would naturally occur at higher orders of the derivation. However, including them in Eq.~\eqref{eq:combination} obtained through considerations up to $\mathcal{O}(\varepsilon^5)$ allows us to conserve the structure of a continuity equation (even with a flux that equals the product of a mobility and a gradient of a potential). We emphasize that this procedure does not correspond to a further approximation. The leading order terms, i.e., terms up to $\mathcal{O}(\varepsilon^5)$ are not touched and adding terms that are smaller than $\mathcal{O}(\varepsilon^5)$ does not change the validity of Eqs.~\eqref{eq:combination}. One may even argue that one has to add these terms to keep the structure as a conservation law intact. See the corresponding discussion for the related problem of a gradient dynamics structure in appendix~A of \cite{ThAP2016prf}.

Furthermore, we introduce the original scales, $\vec x$ and $t$ and the
 fields $\varrho$ and $\varsigma$, e.g.
\begin{equation}
\eta_{\rho \rho} \varepsilon^2 \vec \nabla_{\!\vec X}^2  \left( \varepsilon^2 \rho_1^2 +2 \varepsilon^3  \rho_1 \rho_2 \right) = \eta_{\rho \rho} \varepsilon^2 \vec \nabla_{\!\vec X}^2 \left( \varepsilon \rho_1 + \varepsilon^2 \rho_2\right)^2 + \mathcal{O}(\varepsilon^6)  =  \eta_{\rho \rho}  \vec \nabla_{\!\vec x}^2 \varrho^2 + \mathcal{O}(\varepsilon^6),
\end{equation}
and obtain as result the coupled amplitude equations
\begin{align}\label{eq:general_nonCH_A}
\begin{split}  
\partial_t \varrho = &-\vec \nabla_{\!\vec x} \cdot \left[ \left(Q_0 + q_\rho \varrho + q_\sigma \varsigma + q_{\rho \rho}  \varrho^2 + q_{\rho \sigma} \varrho \varsigma + q_{\sigma \sigma}  \varsigma^2\right) \vec \nabla_{\!\vec x} \left[ \eta_\rho  \varrho +  \eta_\sigma  \varsigma + \eta_{\rho \rho} \varrho^2 + \eta_{\rho \sigma} \varrho \varsigma + \eta_{\sigma \sigma} \varsigma^2  \right. \right. ~\\
& \left. \left. \quad +
\eta_{\rho \rho \rho} \varrho^3 +  \eta_{\rho \rho \sigma} \varrho^2 \varsigma +  \eta_{\rho \sigma \sigma} \varrho \varsigma^2 + \eta_{\sigma \sigma \sigma} \varsigma^3  +  \eta_{\rho xx} \vec \nabla_{\!\vec x}^2 \varrho +\eta_{\sigma xx} \vec \nabla_{\!\vec x}^2 \varsigma
\right] \right]~ \\
\partial_t \varsigma =&-\vec \nabla_{\!\vec x} \cdot \left[ \left(R_0 + r_\rho \varrho + r_\sigma \varsigma + r_{\rho \rho}  \varrho^2 + r_{\rho \sigma} \varrho \varsigma + r_{\sigma \sigma}  \varsigma^2\right) \vec \nabla_{\!\vec x} \left[ \mu_\rho  \varrho +   \mu_\sigma  \varsigma  + \mu_{\rho \rho} \varrho^2 + \mu_{\rho \sigma} \varrho \varsigma + \mu_{\sigma \sigma} \varsigma^2  \right. \right. ~\\
& \left. \left. \quad+
\mu_{\rho \rho \rho} \varrho^3 +  \mu_{\rho \rho \sigma} \varrho^2 \varsigma +  \mu_{\rho \sigma \sigma} \varrho \varsigma^2 + \mu_{\sigma \sigma \sigma} \varsigma^3 +  \mu_{\rho xx} \vec \nabla_{\!\vec x}^2  \varrho +\mu_{\sigma xx} \vec \nabla_{\!\vec x}^2 \varsigma
\right] \right]
\end{split}
\end{align}
where all coefficients are real and well defined through series expansions of $Q, R, \eta, \mu$, and $\vecg{F}$ in Eqs.~\eqref{eq:quant_expand} (see Table~\ref{tab:compare_coeff}). Eq.~\eqref{eq:general_nonCH_A} corresponds to a general nonreciprocal Cahn-Hilliard model. 

By construction the mean values of $\varrho$ and $\varsigma$ vanish, i.e., $\int {\rm d} V \varrho =\int {\rm d} V \varsigma =0$. The presence of the quadratic terms in the potentials indicates that, in analogy to the case of the stationary large-scale instability with conservation law [Eq.~\eqref{eq:CHampl}], subcritical cases can also be captured.
Note that the resulting effective cross-couplings in the second and fourth order terms in Eq.~\eqref{eq:general_nonCH_A} are generic and do not depend on specific preconditions on the studied system. They are ``effective'' as they may arise either directly from any coupling (linear or nonlinear) between the conserved quantities via the local terms within the potentials $\eta$ and $\mu$. Or they can occur due to indirect coupling via couplings to nonconserved fields. 
For example, the cross-coupling in the fourth order terms (that can in the case of constant mobilities be eliminated by a transformation, see below) results from linear couplings between the conserved and nonconserved fields via their second spatial derivatives.
That is, no cross-diffusion is needed in the original Eqs.~\eqref{eq:2Conssystem}. See the definitions of the coefficients in Table~\ref{tab:compare_coeff} and the additional explanations in Section~\ref{sec:app:rd} of the Supplementary Material where we discuss an example. 

Note that all nonreciprocal Cahn-Hilliard models studied in the literature \cite{Zimm1997pa,SaAG2020prx,YoBM2020pnasusa,FrWT2021pre} correspond to particular choices of relations between parameters in the derived general nonreciprocal Cahn-Hilliard model \eqref{eq:general_nonCH_A}. These choices are listed in Table~\ref{tab:compare_models}. Other simplified models may be obtained for systems with specific symmetries in the space of order parameter fields. For instance, if the inversion symmetry $(\varrho,\varsigma)\to(-\varrho,-\varsigma)$ holds, all linear contributions to the mobilities and all quadratic contributions to the nonequilibrium chemical potentials drop out. This is however, only likely to correspond to a generic case if the original conserved fields $\rho$ and $\sigma$ both have the mean value zero. A rotational symmetry in this space would result in a further reduction of the numbers of parameters and could provide another interesting (nongeneric) limiting case. Finally, note that in contrast to part of the nonreciprocal Cahn-Hilliard models in the literature the here derived general model is generic as it does not show the spurious gradient dynamics form analyzed in Ref.~\cite{FHKG2023pre}.

\begin{table}[hbt]
	\begin{tabular}{| c | c | c | c |}
		\hline
		Model &  field variables & zero coefficients & nonzero coefficients~
						\\
				\hline
				\hline~
		\pbox{20cm}{Nonreciprocal Cahn-Hilliard, linear coupling\\  You et al.~\cite{YoBM2020pnasusa} and Frohoff-H\"ulsmann et al.~\cite{FrWT2021pre} ~\\
		$\partial_t \phi_\mu = \vec \nabla^2 \left[- \gamma_\mu \vec \nabla^2 \phi_\mu + \chi_\mu \phi_\mu + \frac13 \phi_\mu^3 + \kappa_{\mu\nu} \phi_\nu \right] $~\\
		with $\frac{1}{|V|}\int_V \mathrm{d} V \phi_\mu =  \phi^0_\mu \quad \mu = A, B \, \, ,\nu\neq \mu$ }
		& \pbox{20cm}{$\varrho = \phi_A - \phi^0_A$~\\
		$\varsigma =\phi_B - \phi^0_B$}    
	&  \pbox{20cm}{$    \eta_{\rho \sigma},\, \eta_{\sigma \sigma}$~\\
		$ \mu_{\rho \rho} , \,  \mu_{\rho \sigma} $~\\
$ \eta_{\rho \rho \sigma},\, \eta_{\rho \sigma \sigma}, \,  \eta_{\sigma \sigma \sigma}$~\\
$ \mu_{\rho \rho \rho} , \, \mu_{\rho \rho \sigma}\, ,\mu_{\rho \sigma \sigma}  $ \\
$\eta_{\sigma xx}$~\\
$ \mu_{\rho xx}$ }
  &  \pbox{20cm}{
  \vspace{0.1cm}$\eta_\rho  = -\chi_A - \left(\phi_A^0\right)^2$~\\ 
  	$\mu_\sigma= -\chi_B- \left(\phi_B^0\right)^2$~\\ 
  $\eta_\sigma  = -\kappa_{AB}$~\\
  $\mu_\rho =-\kappa_{BA}$~\\
   $\eta_{\rho \rho}=-\phi_A^0$~\\
   $\mu_{\sigma \sigma} =-\phi_B^0$~\\
   $\eta_{\rho \rho \rho}=\mu_{\sigma\sigma\sigma}=-\frac13$~\\
   $\eta_{\rho xx} = \gamma_A$~\\
$\mu_{\sigma xx} = \gamma_B$\vspace{0.1cm}}   
  \\
				\hline
		\pbox{20cm}{Nonreciprocal Cahn-Hilliard, nonlinear reciprocal,\\ and linear nonreciprocal coupling: Saha et al.~\cite{SaAG2020prx} ~\\
		  $\partial_t \phi_1 = \vec \nabla^2 \bigg[-\kappa \vec \nabla^2 \phi_1 + 2(c_{1,1}+ c_{1,2})^2 \phi_1 - 6 ( c_{1,1} + c_{1,2})\phi_1^2 $~\\
		  $\qquad +4 \phi_1^3 + (\chi + \alpha) \phi_2 + 2 \chi' \phi_1 \phi_2^2 \bigg]$ ~\\
		 $\partial_t \phi_2 = \vec \nabla^2 \bigg[- \kappa \vec \nabla^2 \phi_2 +2(c_{2,1}+ c_{2,2})^2 \phi_2 - 6 ( c_{2,1} + c_{2,2})\phi_2^2$~\\
		 $ \qquad  +4 \phi_2^3 +  (\chi - \alpha) \phi_1+ 2 \chi' \phi_2 \phi_1^2 \bigg]$~\\
		 with $\frac{1}{|V|}\int_V \mathrm{d} V \phi_i = \bar \phi_i \quad i=1,2$}	&
	 \pbox{20cm}{$\varrho = \phi_1 -\bar \phi_1$~\\
		 $\varsigma =\phi_2 -\bar \phi_2$}    
	 &  \pbox{20cm}{	
	 $ \eta_{\rho \rho \sigma},\,  \eta_{\sigma \sigma \sigma}$~\\
	 $ \mu_{\rho \rho \rho} ,\, \mu_{\rho \sigma \sigma}  $ \\
	 $\eta_{\sigma xx}$~\\
	 $ \mu_{\rho xx}$ }
 &  \pbox{20cm}{
 \vspace{0.1cm}
 $\eta_\rho  = -2(c_{1,1}+ c_{1,2})^2 - 12 \bar \phi_1^2 - 2 \chi' \bar \phi_2^2$~\\ 
 $\mu_\sigma= -2(c_{2,1}+ c_{2,2})^2 - 12 \bar \phi_2^2- 2 \chi' \bar \phi_1^2$~\\ 
 $\eta_\sigma  = -\chi - \alpha - 4 \chi' \bar \phi_1 \bar \phi_2 $~\\
 $\mu_\rho = -\chi + \alpha- 4 \chi' \bar \phi_1 \bar \phi_2$~\\
 $\eta_{\rho \rho}=6 ( c_{1,1} + c_{1,2}) - 12 \bar \phi_1$~\\
 $\mu_{\sigma \sigma} =6 ( c_{2,1} + c_{2,2}) -12 \bar \phi_2$~\\
 $\eta_{\sigma \sigma}= -2 \chi' \bar \phi_1$~\\
 $\eta_{\rho \sigma}= -4 \chi' \bar \phi_2$~\\ 
 $ \mu_{\rho \rho}= -2 \chi' \bar \phi_2$~\\
 $  \mu_{\rho \sigma}=-4 \chi' \bar \phi_1 $~\\
 $\eta_{\rho \sigma \sigma}= \mu_{\rho \rho \sigma}= -2 \chi'$~\\
 $\eta_{\rho \rho \rho}=\mu_{\sigma\sigma\sigma}=-4$~\\
 $\eta_{\rho xx} = \mu_{\sigma xx} = \kappa$
 \vspace{0.1cm}}   ~\\
		 		\hline
          \pbox{20cm}{Complex Cahn-Hilliard\\
          Zimmermann  \cite{Zimm1997pa}~\\ $\partial_t A =  -G \vec \nabla^2 \left[ \varepsilon + (1 + \mathrm{i}b) \vec \nabla^2 -(1 + \mathrm{i} c) |A|^2 \right]A$ ~\\
		  with $A= A_r + \mathrm{i} A_i$}
		  	&  \pbox{20cm}{$\varrho = A_r$~\\
		 $\varsigma =A_i$}  
		  &  \pbox{20cm}{$\eta_\sigma  , \, \mu_\rho $~\\ 
		  $\eta_{\rho \rho} , \,  \eta_{\rho \sigma},\, \eta_{\sigma \sigma}$~\\
		$ \mu_{\rho \rho} , \,  \mu_{\rho \sigma}  , \,  \mu_{\sigma \sigma} $~}
				  & \pbox{20cm}{
				  \vspace{0.1cm}
				  $\eta_\rho =\mu_\sigma = G \varepsilon$ ~\\
		  $\eta_{\rho \sigma \sigma}=\eta_{\rho \rho \rho}=\mu_{\rho \rho \sigma}=\mu_{\sigma \sigma \sigma}= -G$~\\
		  $\mu_{\rho \sigma \sigma}=\mu_{\rho \rho \rho}=-\eta_{\rho \rho \sigma}=-\eta_{\sigma \sigma \sigma}= Gc$~\\
		  $\eta_{\sigma xx} =- \mu_{\rho xx} =-G b$~\\
		  $\eta_{\rho xx} = \mu_{\sigma xx} = G$\vspace{0.1cm}}~\\
		  		\hline
		  \pbox{20cm}{ Reciprocal Cahn-Hilliard [variational structure] ~\\
		   $\partial \phi_i = \vec \nabla^2 \frac{\delta \mathcal{F}}{\delta \phi_i} $~\\
		   with  $\mathcal{F}=\sum_i \mathcal{F}_i+\mathcal{F}_\text{coup}$~\\
		   $\mathcal{F}_i= \sum_i \int \mathrm{d} V  \left(\frac{\kappa_i}{2} (\vec \nabla \phi_i)^2 + \frac{\alpha_i}{2} \phi_i^2 +\frac{\beta_i}{3} \phi_i^3+ \frac{\gamma_i}{4} \phi_i^4\right)$ ~\\
		   $\mathcal{F}_\text{coup} = \int \mathrm{d} V\bigg[ K  \vec \nabla \phi_1 \vec \nabla \phi_2 +  a \phi_1 \phi_2 + b_1 \phi_1^2 \phi_2 + b_2 \phi_1 \phi_2^2 $~\\
		   $\qquad + c_1 \phi_1^3 \phi_2+ c_2 \phi_1^2 \phi_2^2 + c_3 \phi_1 \phi_2^3\bigg] $~\\
	   $ \frac{1}{|V|}\int_V\mathrm{d} V \phi_i =0$}	&\pbox{20cm}{$\varrho = \phi_1$~\\
		   $\varsigma =\phi_2$}    &  -- & 
		   \pbox{20cm}{ 
		   \vspace{0.1cm}
		   $\eta_\rho  = -\alpha_1$~\\ 
		   	$\mu_\sigma= -\alpha_2$~\\ 
		   	$\eta_{\rho \rho}= -\beta_1$~\\
		   	$\mu_{\sigma \sigma}= -\beta_2$~\\
		   	$\eta_{\rho \rho \rho}= -\gamma_1$~\\
		   	$\mu_{\sigma \sigma \sigma}= -\gamma_2$~\\
		   	$\eta_{\rho xx}= \kappa_1$~\\
		   	$\mu_{\sigma xx}= \kappa_2$~\\
		   	 $\eta_\sigma= \mu_\rho = -a$~\\
		   $ \eta_{\sigma xx}= \mu_{\rho xx}=K$~\\
		   $\eta_{\rho \sigma}= 2 \mu_{\rho \rho}=-2b_1$~\\
		   $ 2 \eta_{\sigma \sigma}= \mu_{\rho \sigma}=-2b_2$~\\
		   $\eta_{\rho \rho \sigma}= 3 \mu_{\rho \rho \rho}=-3 c_1$~\\
		   $\eta_{\rho \sigma \sigma}= \mu_{\rho \rho \sigma}=-2 c_2$~\\
		   $3 \eta_{\sigma \sigma \sigma}= \mu_{\rho \sigma \sigma}=-3 c_3$\vspace{0.1cm}
		   } ~\\
		\hline
	\end{tabular}
	\caption{Identification of models studied in the literature \cite{Zimm1997pa,SaAG2020prx,YoBM2020pnasusa,FrWT2021pre} as special cases of the here derived general nonreciprocal Cahn-Hilliard model in the form of Eq.~\eqref{eq:general_nonCH_A}.  All listed models only consider constant mobilities $Q=Q_0=1$ and $R=R_0=1$, i.e., $q_{\rho}=q_{\sigma}=q_{\rho\rho}=q_{\rho\sigma}=q_{\sigma\sigma}=0$ and $r_{\rho}=r_{\sigma}=r_{\rho\rho}=r_{\rho\sigma}=r_{\sigma\sigma}=0$. Furthermore the relation to the reciprocal limiting case is given.}
	\label{tab:compare_models}
\end{table}

\FloatBarrier

In matrix form Eq.~\eqref{eq:general_nonCH_A} reads
	\begin{align}	 \label{eq:general_nonCH_matrix}
\partial_t \left(\begin{array}{c}
\varrho\\
\varsigma
\end{array}\right)
= -\vec \nabla  \cdot \left[ \tens{M}(\varrho,\varsigma)
\vec \nabla \left( 
\tens{L}		\left(\begin{array}{c}
\varrho\\
\varsigma
\end{array}\right)+ 
\left(\begin{array}{c}
N_\rho(\varrho,\varsigma)\\
N_\sigma(\varrho,\varsigma)
\end{array}\right)
+
\tens{D}
\vec \nabla^2
\left(\begin{array}{c}
\varrho\\
\varsigma
\end{array}\right)
\right) 
\right]
\end{align}
where $\tens{M}(\varrho,\varsigma)$ is a non-constant diagonal mobility matrix, and $\tens{L}$ and $\tens D$ are fully occupied constant matrices that describe the terms within the potentials that are linear in the fields and linear in their second spatial derivatives, respectively. All nonlinearities within the potentials are contained in the general cubic functions $N_\rho(\varrho,\varsigma)$ and $N_\sigma (\varrho,\varsigma)$.
The matrix form clearly shows that the system of coupled nonlinear equations features linear and nonlinear second order and linear fourth order cross-coupling terms (see discussion at Eqs.~\eqref{eq:general_nonCH_A}). One may distinguish dynamic cross-coupling (via $\tens{M}$) and energetic cross-coupling (via $\tens{L}$, $\tens{D}$ and the nonlinearities).

If we linearize Eqs.~\eqref{eq:general_nonCH_matrix}  we obtain
\begin{align}\label{eq:lin_amplitudeEq}
\partial_t \left(\begin{array}{c}
\varrho\\
\varsigma
\end{array}\right)
=& -\vec \nabla\cdot\left[ \tens{M}(0,0)
\vec \nabla \left( 
\tens{L}		\left(\begin{array}{c}
\varrho\\
\varsigma
\end{array}\right)+ 
\tens{D}
\vec \nabla^2
\left(\begin{array}{c}
\varrho\\
\varsigma
\end{array}\right)
\right) 
\right],~\nonumber \\
\mathrm{i.e.,}\qquad
	\partial_t \left(\begin{array}{c}
		\varrho\\
		\varsigma
	\end{array}\right)
	=& -\left(\begin{array}{c c}
		Q_0\eta_\rho  & 	Q_0 \eta_\sigma ~\\
			R_0\mu_\rho  & 	R_0 \mu_\sigma 
	\end{array}\right)
	\vec \nabla^2 
	\left(\begin{array}{c}
		\varrho\\
		\varsigma
			\end{array}\right)
		-
		\left(\begin{array}{c c}
			Q_0 \eta_{\rho xx} & 	Q_0\eta_{\sigma xx}~\\
			R_0\mu_{\rho xx} & 	R_0\mu_{\sigma xx}
		\end{array}\right)
\vec \nabla^4 \left(\begin{array}{c}
			\varrho\\
			\varsigma
	\end{array}\right).
\end{align}
Determining the eigenvalues we recover the original dispersion relation~(see Eq.~(1) of the main text) as it should be. In particular, the coefficients in Eq.~\eqref{eq:lin_amplitudeEq} are related to the ones of Eq.~(1) of the main text as
 \begin{align}
\tilde \delta =& 		-\frac{Q_0 \eta_{\rho xx} + 	R_0\mu_{\sigma xx}}{2}~\\
 \mathrm{i}\tilde \omega=& 	\frac{\eta_\rho \eta_{\rho xx} Q_0^2 + (2 \eta_{\sigma xx} \mu_{\rho} + 2 \eta_{\sigma} \mu_{\rho xx} - \eta_{\rho xx} \mu_{\sigma} - 
 	\eta_{\rho} \mu_{\sigma xx}) Q_0 R_0 + \mu_{\sigma} \mu_{\sigma xx} R_0^2}{2\sqrt{
 	\eta_\rho^2 Q_0^2 - 2 \eta_\rho \mu_\sigma Q_0 R_0 + R_0 (4 \eta_\sigma \mu_\rho Q_0 + \mu_\sigma^2 R_0)}}
 	\end{align}
 together with the already known relations for $\delta_2$ and $\omega$ from Eqs.~\eqref{eq:delta_omega}. Note that in contrast to the dynamics on the timescale $\tau$, here, we take the deviation from the onset of instability into account, i.e., $(Q_0 \eta_\rho  + R_0 \mu_\sigma)/2  = \delta_2 \varepsilon^2$.

Finally, we may remove certain terms by formulating Eqs.~\eqref{eq:general_nonCH_matrix} in alternative conserved amplitudes $ \left(\begin{array}{c}
	A\\
	B
	\end{array}\right)\equiv \tens{T}\,\left(\begin{array}{c}
	\varrho \\
	\varsigma
	\end{array}\right)$ where $\tens{T}$ is the corresponding transformation matrix. 
	Multiplying Eqs.~\eqref{eq:general_nonCH_matrix} with $\tens{T}$ we rewrite it in the new field variables as
		\begin{align}	 \label{eq:general_nonCH_matrix-two}
	\partial_t \left(\begin{array}{c}
	A\\
	B
	\end{array}\right)
	= -\vec \nabla \cdot  \left[\tens{T}\, \tens{\widetilde M}(A,B) \tens{T}^{-1}
	\vec \nabla\left( 
	\tens{T}\,\tens{L}\, \tens{T}^{-1}		\left(\begin{array}{c}
	A\\
	B
	\end{array}\right)+
	\tens{T} 
	\left(\begin{array}{c}
	\widetilde N_\rho(A,B)\\
	\widetilde N_\sigma(A,B)
	\end{array}\right)
	+
	\tens{T}\,
	\tens{D}\,
	\tens{T}^{-1}
	\vec \nabla^2
	\left(\begin{array}{c}
	A\\
	B
	\end{array}\right)
	\right) 
	\right]
\end{align}
where the quantities with tilde are obtained by, e.g., $\tens{M}(\varrho,\varsigma)=\tens{M}(\varrho(A,B),\varsigma(A,B))=\tens{\widetilde M}(A,B)$.

If we employ a transformation matrix $\tens T$ such that $\tens{T}\, \tens{D}\, \tens{T}^{-1}$ is diagonal we can eliminate the linear (energetic) cross-coupling in the fourth order derivative terms. However, in general, the resulting mobility matrix $\tens{T}\,\tens{\widetilde M}\,\tens{T}^{-1}$ will not be diagonal, i.e., one replaces the energetic cross-coupling by a dynamic one. This is different if the mobility matrix is constant, i.e., $\tens M=\tens M_0$. Then, $\tens M_0$ can be moved behind the gradient operator allowing one to employ a transformation matrix $\tens{T}$ that diagonalizes the product $ \tens{M}_0\, \tens{D}$. In this case, the cross-coupling in the highest order terms can be eliminated completely. The resulting equation is
 		\begin{align}	 \label{eq:general_nonCH_matrix-three}
         \partial_t \left(\begin{array}{c}
	A\\
	B
	\end{array}\right)
	= \vec \nabla^2 \left[
	{\tens{\widetilde L}}		\left(\begin{array}{c}
	A\\
	B	\end{array}\right)+
	\left(\begin{array}{c}
	N_A(A,B)\\
	N_B(A,B)
	\end{array}\right)
	-
		\left(\begin{array}{c c}
			D_A & 	0~\\
			0 & 	D_B
		\end{array}\right)
	\vec \nabla^2
	\left(\begin{array}{c}
	A\\
	B
	\end{array}\right)
	\right]
                \end{align}
where $\tens{\widetilde L}=-\tens{T}\, \tens{M}_0 \,\tens{L}\, \tens{T}^{-1}$ and $(N_A,N_B)^T=-\tens{T}\, \tens{M}_0 (\widetilde N_\rho, \widetilde N_\sigma)^T$. 
These are the coupled equations given in the main text.

\begin{figure}[bth]
	\includegraphics[width=0.7\columnwidth]{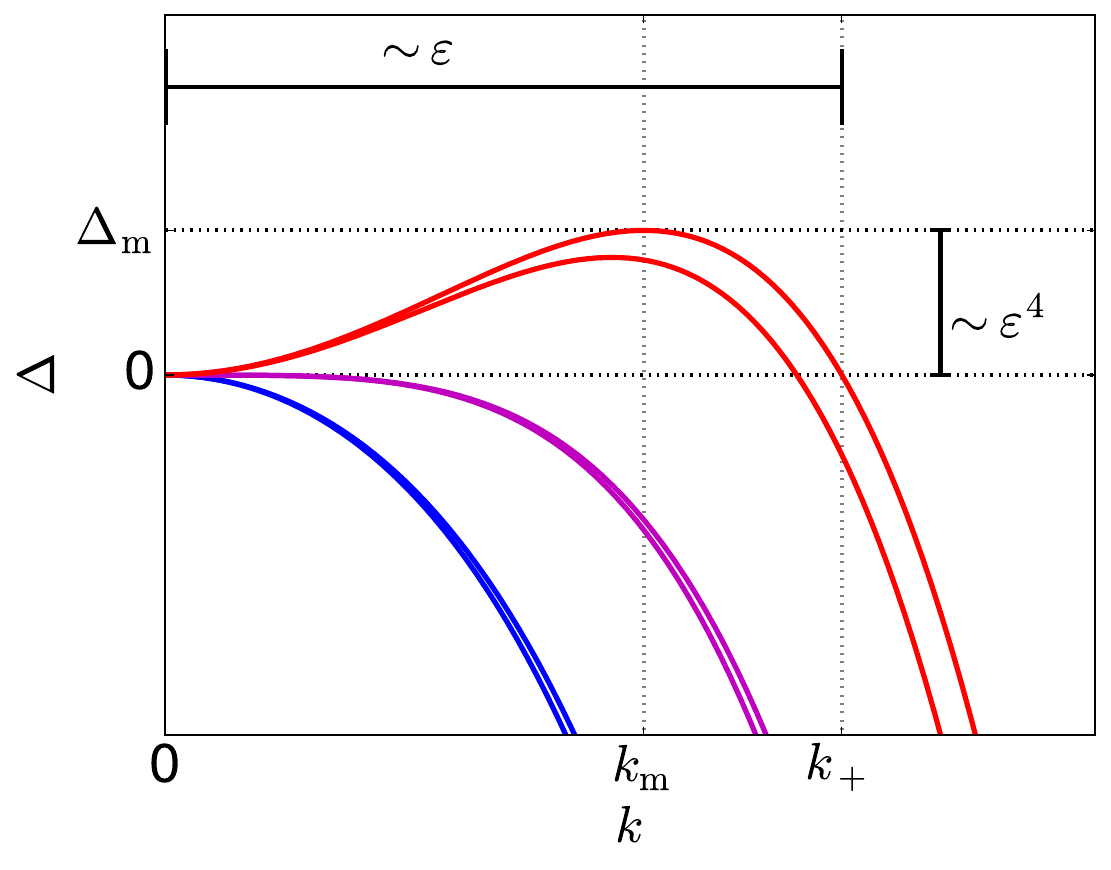}
	\caption{Dispersion relations \eqref{eq:lambda2CH} below (blue lines), at (purple line) and above (red line) the threshold of a codimension-2 Cahn-Hilliard instability where two Cahn-Hilliard modes simultaneously become unstable. Labeled dotted lines and solid bars indicate typical quantities and scalings above onset.}
	\label{fig:onset_2CH}
\end{figure}

\subsection{Amplitude equation for codimension-2 Cahn-Hilliard instability}\label{sec:2CH_AE}
Up to here we have considered the conserved-Hopf instability as one of eight basic codimension-1 instabilities and have derived the nonreciprocal Cahn-Hilliard model as a corresponding generic amplitude equation. Next, we furthermore show that they
also correspond to the amplitude equation for a stationary large-scale codimension-2 instability that involves two conservation laws, i.e., for two simultaneously occurring Cahn-Hilliard instabilities.
  
In particular, we consider the dispersion relation for two Cahn-Hilliard modes, i.e., characterized by two real eigenvalues given by
\begin{equation}\label{eq:lambda2CH}
\lambda_{\pm}(k)= \delta_\pm k^2 - \tilde \delta_\pm k^4 + \mathcal{O}(k^6)\,.
\end{equation}
Considering the codimension-2 instability where both real modes become simultaneously unstable at control parameter $\varepsilon=0$, both leading order growth rates are small, i.e., $\delta_\pm = \delta_{\pm, 2} \varepsilon^2$. Fig.~\ref{fig:onset_2CH} illustrates the dispersion relations below, at and above the onset. The indicated scalings of the band of unstable wavenumbers and of the maximal growth rate are as in Fig.~1 of the main text.

Again we consider the general multi-component model~\eqref{eq:2Conssystem}, use the same ansatz \eqref{eq:fields_expand} and only discuss the differences for the current case as compared to the analysis done in Section~\ref{sec:app:consHopf} of the Supplementary Material.
First, all terms at order $\varepsilon$ and $\varepsilon^2$ are unchanged. We also find the same linear system \eqref{eq:leading_osc0} at order $\varepsilon^3$. However, in contrast to the case of the conserved-Hopf instability, the resulting eigenvalues are real, i.e., the inequality~\eqref{eq:inequality} is reversed and we identify
\begin{equation}
 \frac{Q_0 \eta_\rho + R_0 \mu_\sigma}{2}   \pm  \sqrt{\frac{(Q_0 \eta_\rho -R_0 \mu_\sigma)^2}{4} + Q_0 \eta_\sigma R_0 \mu_\rho  }=\delta_\pm \,.
\end{equation}
Now we demand that both $\delta_+$ and $\delta_-$ are $\mathcal{O}(\varepsilon^2)$ and, thus, we can set them to zero at this order. In consequence, there is no dynamics on timescale $\tau$ which is as expected since there is no oscillation in contrast to the conserved-Hopf case. In other words, while for the case of a conserved-Hopf instability the leading order frequency is $\omega k^2$ with $\omega = \mathcal{O}(1)$, here $\omega$ is $\mathcal{O}(\varepsilon^2)$ and imaginary, i.e., is simply part of the growth rate.

Proceeding to orders $\varepsilon^4$ and $\varepsilon^5$, equations~\eqref{eq:order4}-\eqref{eq:order5_2} are recovered and using the same recombination as in Eq.~\eqref{eq:combination}, again we obtain amplitude equations that correspond to the generalized nonreciprocal Cahn-Hilliard model~\eqref{eq:general_nonCH_A}. Linearization yields Eq.~\eqref{eq:lin_amplitudeEq}, and calculating the eigenvalues, again we recover the original dispersion relation, i.e., we identify
\begin{equation}
 		\frac{Q_0 \eta_{\rho xx} + 	R_0\mu_{\sigma xx}}{2} \pm
 	\frac{\eta_\rho \eta_{\rho xx} Q_0^2 + (2 \eta_{\sigma xx} \mu_{\rho} + 2 \eta_{\sigma} \mu_{\rho xx} - \eta_{\rho xx} \mu_{\sigma} - 
 	\eta_{\rho} \mu_{\sigma xx}) Q_0 R_0 + \mu_{\sigma} \mu_{\sigma xx} R_0^2}{2\sqrt{
 	\eta_\rho^2 Q_0^2 - 2 \eta_\rho \mu_\sigma Q_0 R_0 + R_0 (4 \eta_\sigma \mu_\rho Q_0 + \mu_\sigma^2 R_0)}}= -\tilde{\delta}_\pm\,.
\end{equation} 
We conclude that beside the conserved-Hopf instability the generalized nonreciprocal Cahn-Hilliard model also describes the generic behavior close to the simultaneous onset of two large-scale stationary instabilities with conservation laws, i.e., two Cahn-Hilliard instabilities. It is valid when the inequality $\eqref{eq:inequality}$ is reversed, i.e., when the imaginary part of the eigenvalues vanishes, and the resulting additional contributions to the growth rates are still small, i.e., $\mathcal{O}(\varepsilon^2)$.

Note that, in general, there are no further emerging conditions on the coefficients of the nonlinear terms in Eq.~\eqref{eq:general_nonCH_A}, i.e., despite the stationary character of the considered codimension-2 instability, the resulting amplitude equations are normally still nonreciprocal, i.e., nonvariational, if the original model is nonvariational. In practice, this implies that nonlinear states resulting from secondary, tertiary, etc.\ instabilities may exhibit time-dependent (periodic or irregular) behavior. In contrast, if the original model is itself variational the resulting amplitude equations will directly inherit the variational structure. i.e., the parameters of the nonreciprocal Cahn-Hilliard model acquire mutual relations that render it a reciprocal Cahn-Hilliard model. The corresponding conditions for the various parameters are given in the final row of Table~\ref{tab:compare_models}. This occurs, for instance, in the case of dewetting isothermal two-layer liquid films on solid substrates \cite{PBMT2005jcp,JPMW2014jem} and decomposing ternary mixtures \cite{MKHK2019sm,Ma2000jpsj}.

The presented derivation indicates that the generalized nonreciprocal Cahn-Hilliard model is not only an amplitude equation for the conserved-Hopf instability but represents an amplitude equations for a bifurcation of higher codimension. In other words, it belongs to a higher level of a hierarchy of amplitude equations obtained in the vicinity of bifurcations of successively higher codimension. One may say that the eight codimension-1 cases discussed in Section~\ref{sec:app:disp-amp} of the Supplementary Material form the base layer of the hierarchy. The two derivations of the nonreciprocal Cahn-Hilliard model presented in the Supplementary Material imply that it is (at least) the amplitude equation for the higher codimension point where conserved-Hopf instability and the codimension-2 Cahn-Hilliard instability transform into each other. This, in hindsight explains why the nonreciprocal Cahn-Hilliard models analyzed in Refs.~\cite{YoBM2020pnasusa,SaAG2020prx,FrWT2021pre,FrTh2021ijam} show such a rich and varied behavior. Note that a further reduction to an amplitude equation that only describes the conserved-Hopf instability is a nontrivial technical challenge and remains a task for the future. Such a reduced equation would feature fewer parameters, solely focus on the behavior that is universal for the conserved-Hopf instability and not include any other instability. Note that the relations between amplitude equations on different hierarchy level is not trivial~\cite{RoDo1998pd}. 

\subsection{Example: Three-species reaction-diffusion system with two conservation laws}
  \label{sec:app:rd}
Here, we exemplarily derive the amplitude equation for a
  specific system and check the agreement of its solutions with the
  solutions of the original system. In particular, we use a
  three-component reaction diffusion system with two local
  conservation laws. Its general form is
\begin{align}
\begin{split}\label{eq:RD3}
\partial_t u =&  D_u  \vec \nabla^2 u + \alpha f(u,v,w),~\\
\partial_t v =&  D_v  \vec \nabla^2 v + \beta f(u,v,w),~\\
\partial_t w = & D_w  \vec \nabla^2 w + \gamma f(u,v,w).
\end{split}
\end{align}
Here, we use the cubic expression $f(u, v, w) = a u + b v + c w + d(u - v)^2 - w^3$.
The stability of the homogeneous steady state $(u_0,v_0,w_0)=(0,0,0)$ is determined by the Jacobian
\begin{equation}
\tens{J}(k^2)= 
\left(
\begin{array}{c c c}
-D_u k^2 + \alpha a & \alpha b & \alpha c~\\
\beta a & - D_v k^2 + \beta b & \beta c ~\\
\gamma a & \gamma b & -D_w k^2 + \gamma c
\end{array}
\right)
\end{equation}
that exhibits two neutral modes $(-c, 0, a)$ and $(-b, a, 0)$ for $k=0$. Applying a Taylor expansion we can write the corresponding two dispersion relations as Eq.~(1) of the main text with
\begin{align}
\begin{split}
\Delta(k) = &-\frac{k^2}{2(a \alpha + b \beta + c \gamma)}
\left(a (D_v + D_w) \alpha + b (D_u+ D_w) \beta + c (D_u + D_v) \gamma \right) + \mathcal{O}(k^4),\\
\Omega(k)=&\frac{k^2
	\sqrt{4 (a \alpha + b \beta + c \gamma) (a D_v D_w \alpha + 
		b D_u D_w \beta + c D_u D_v \gamma) - (a (D_v + D_w) \alpha + 
		b (D_u + D_w) \beta + c (D_u + D_v) \gamma)^2}}{2(a \alpha + b \beta + c \gamma)} + \mathcal{O}(k^4).
\end{split}
\end{align}
We identify the critical value for $a$ as
\begin{equation}\label{eq:ac}
a_\text{c}=-\frac{b( D_u+D_w) \beta + c (D_u+D_v)\gamma}{(D_v + D_w) \alpha},
\end{equation}
and the leading order frequency is given by
\begin{align}
\Omega(k)=&\omega k^2 + \mathcal{O}(k^4)~\\
\text{with}\,\,\, \omega=&
\sqrt{\frac{b \beta  D_w^2
		(D_u-D_v)+c \gamma  D_v^2 (D_u-D_w)}{b \beta 
	(D_v-D_u)+c \gamma  (D_w-D_u)}}.
\end{align}
This defines the onset of the conserved-Hopf instability. In the
following we set $a=a_\text{c} + \varepsilon^2 a_2$.
The conserved quantities are $\rho=-\gamma u + \alpha w$ and
$\sigma=-\beta u + \alpha v$. Note that any linear combination for
both, the neutral modes and the conserved quantities, result in equivalent descriptions.
Rewriting Eqs.~\eqref{eq:RD3} in the two conserved quantities $\rho$, $\sigma$ and simply using $n=\alpha w$ as the remaining nonconserved quantity gives
\begin{align}
\begin{split}\label{eq:RD3_rewritten}
\partial_t \rho =&   \vec \nabla^2\left( D_u \rho  + (D_w- D_u)n \right)~\\
\partial_t \sigma =&    \vec \nabla^2\left(  D_v \sigma + \frac{\beta}{\gamma} (D_u - D_v)(\rho - n) \right)~\\
\partial_t n = &   D_w \vec \nabla^2 n+  (a \alpha + b \beta+c \gamma ) n - (\alpha a + \beta b) \rho + \gamma  b \sigma + \frac{\left( (\beta - \alpha)n+ (\alpha - \beta) \rho + \gamma \sigma \right)^2}{\alpha \gamma} - \frac{\gamma}{\alpha^2} n^3
\end{split}
\end{align}
Eqs.~\eqref{eq:RD3_rewritten} is a special case of Eqs.~\eqref{eq:2Conssystem} characterized by constant mobilities, i.e., $Q=R=1$, and by potentials $\eta$, $\mu$ that are linear in the order parameters. Note that Eqs.~\eqref{eq:RD3_rewritten} do not contain any fourth order derivatives.

Now, performing the weakly nonlinear analysis as explained in Section~\ref{sec:app:consHopf} of the Supplementary Material, we can relate all coefficients of the original example model~\eqref{eq:RD3_rewritten} with the coefficients that occur in the course of the calculation in Section~\ref{sec:app:consHopf} and in the finally obtained amplitude equation~\eqref{eq:general_nonCH_A}. First, the coefficients that determine the nonconserved field $n$ as a (nonlinear) function of the conserved quantities $\rho$ and $\sigma$ at the different orders in $\varepsilon$ are
\begin{align}
\begin{split}\label{eq:coeff}
\text{Eq.}~\eqref{eq:n1}:& \quad n_\rho=  \frac{\alpha a + \beta b}{a \alpha + b \beta+c \gamma }, \qquad
n_\sigma= -\frac{\gamma  b}{a \alpha + b \beta+c \gamma },~\\
\text{Eq.}~\eqref{eq:n2}:& \quad n_{\rho \rho}= -\frac{c^2 \gamma  d (\alpha -\beta
   )^2}{\alpha  (a \alpha +b \beta +c \gamma
   )^3}, \qquad
n_{\rho \sigma}= -\frac{2 c \gamma  d (\alpha -\beta ) (a
   \alpha +\alpha  b+c \gamma )}{\alpha  (a
   \alpha +b \beta +c \gamma )^3}, \qquad
n_{\sigma \sigma}= -\frac{\gamma  d (a \alpha +\alpha  b+c \gamma
   )^2}{\alpha  (a \alpha +b \beta +c \gamma
   )^3},\\
\text{Eq.}~\eqref{eq:n3}:& \quad n_{\rho \rho \rho}=-\frac{\gamma  \left(2 c^3 d^2 (\alpha -\beta
   )^4-(a \alpha +b \beta )^3 (a \alpha +b
   \beta +c \gamma )\right)}{\alpha ^2 (a
   \alpha +b \beta +c \gamma )^5},~\\
& \quad  n_{\rho \rho \sigma}= 6-\frac{3 \gamma  \left(2 c^2 d^2 (\alpha
   -\beta )^3 (\alpha  (a+b)+c \gamma )+b
   \gamma  (a \alpha +b \beta )^2 (a \alpha +b
   \beta +c \gamma )\right)}{\alpha ^2 (a
   \alpha +b \beta +c \gamma )^5},
~\\
& \quad n_{\rho \sigma \sigma}= 6-\frac{3 \gamma  \left(2 c d^2 (\alpha -\beta
   )^2 (\alpha  (a+b)+c \gamma )^2-b^2 \gamma
   ^2 (a \alpha +b \beta ) (a \alpha +b \beta
   +c \gamma )\right)}{\alpha ^2 (a \alpha +b
   \beta +c \gamma )^5},
	 ~\\
&	\quad  n_{\sigma \sigma \sigma} =  -\frac{\gamma  \left(b^3 \gamma ^3 (a \alpha
   +b \beta +c \gamma )+2 d^2 (\alpha -\beta )
   (\alpha  (a+b)+c \gamma )^3\right)}{\alpha
   ^2 (a \alpha +b \beta +c \gamma )^5},\\
& \quad  n_{\rho xx}= \frac{c \gamma  (a \alpha 
   (D_u-D_w)+b \beta 
   (D_v-D_w))}{(a \alpha +b \beta
   +c \gamma )^3}, \qquad
n_{\sigma xx}= \frac{b \gamma  (a \alpha 
   (D_u-D_v)+c \gamma 
   (D_w-D_v))}{(a \alpha +b \beta
   +c \gamma )^3}.
\end{split}
\end{align} 
For our specific example the continuity equations (the first two equations of~\eqref{eq:RD3_rewritten}) are purely linear. Therefore, all nonlinear terms in the final amplitude equation~\eqref{eq:general_nonCH_A} are a direct result of the coupling to the nonconserved field. Namely, the coefficients in \eqref{eq:general_nonCH_A} result as
\begin{align}
\begin{split}\label{eq:coeff_2}
\eta_\rho=&(D_u-D_w) n_\rho -D_u \, , \qquad
\eta_\sigma=(D_u-D_w) n_\sigma\, ,~~\\
   \eta_{\rho \rho} = &(D_u-D_w) n_{\rho \rho}\, , \qquad \eta_{\rho \sigma} = (D_u-D_w) n_{\rho \sigma}\, , \qquad \eta_{\sigma \sigma} = (D_u-D_w) n_{\sigma \sigma}\,,~\\   
   \eta_{\rho \rho \rho} = &(D_u-D_w) n_{\rho \rho \rho}\, , \qquad \eta_{\rho \rho \sigma} = (D_u-D_w) n_{\rho \rho \sigma}\, , \qquad \eta_{\rho \sigma \sigma} = (D_u-D_w) n_{\rho \sigma \sigma} \, , \qquad \eta_{\sigma \sigma \sigma} = (D_u-D_w) n_{\sigma \sigma \sigma}\, ,~\\
   \eta_{\rho xx} = &(D_u-D_w) n_{\rho xx}\, , \qquad \eta_{\sigma xx} = (D_u-D_w) n_{\sigma xx}\, ,\\  
\mu_\rho=&  \frac{\beta}{\gamma}(D_u-D_v)(n_\rho -1)\, , \qquad
\mu_\sigma= \frac{\beta}{\gamma}(D_u-D_v)n_\sigma-D_v\,,~~\\
   \mu_{\rho \rho} = &\frac{\beta}{\gamma} (D_u - D_v) n_{\rho \rho}\, , \qquad \mu_{\rho \sigma} = \frac{\beta}{\gamma} (D_u - D_v) n_{\rho \sigma}\, , \qquad \mu_{\sigma \sigma} = \frac{\beta}{\gamma} (D_u - D_v) n_{\sigma \sigma}\,,~\\   
   \mu_{\rho \rho \rho} = &\frac{\beta}{\gamma} (D_u - D_v) n_{\rho \rho \rho}\, , \qquad \mu_{\rho \rho \sigma} = \frac{\beta}{\gamma} (D_u - D_v) n_{\rho \rho \sigma}\, , \qquad \mu_{\rho \sigma \sigma} = \frac{\beta}{\gamma} (D_u - D_v) n_{\rho \sigma \sigma} \, , \qquad \mu_{\sigma \sigma \sigma} = \frac{\beta}{\gamma} (D_u - D_v) n_{\sigma \sigma \sigma}\, ,~\\
      \mu_{\rho xx} = &\frac{\beta}{\gamma} (D_u - D_v) n_{\rho xx}\, , \qquad \mu_{\sigma xx} = \frac{\beta}{\gamma} (D_u - D_v) n_{\sigma xx}\, .
   \end{split}
\end{align}
Since the mobilities in the original model equations are constant ($Q_0=R_0=1$) all coefficients $q_i$ and $r_i$ in Eq.~\eqref{eq:general_nonCH_A} are identical zero.
Note that in the amplitude equation all fourth order derivatives occur including the two cross-coupling terms represented by $\eta_{\sigma xx}$ and $\mu_{\rho xx}$, respectively. This occurs even though fourth order derivatives are absent in the original model~\eqref{eq:RD3_rewritten}. Here, they arise as an effective coupling via the nonconserved field similar to the nonlinear terms.
Finally, note that the parameter $a$ occurs in all
coefficients~\eqref{eq:coeff}, although its critical value would
mostly be sufficient to leading order. Only in the coefficients of the
linear terms, i.e., $\eta_\rho$, $\eta_\sigma$, $\mu_\rho$, and
$\mu_\sigma$ the smallness parameter $\varepsilon$ has to be
considered. However, for simplicity and without the need of making any further approximations, we can equally use the exact value $a$. See the main text for a similar discussion.

Finally, we check the validity of the derived amplitude equation. To
do so we perform numerical path continuation \cite{DWCD2014ccp,UeWR2014nmma} on the one hand for the
original reaction-diffusion system~\eqref{eq:RD3_rewritten} and on the
other hand for the derived amplitude
equation~\eqref{eq:general_nonCH_A} employing the coefficients of
Eq.~\eqref{eq:coeff}. Finally, we compare the results.
We use $a$ as control parameter and consider the deviation from its critical value $a_\text{c}$ (Eq.~\eqref{eq:ac}) that defines the onset of the conserved-Hopf instability. For the chosen parameters the trivial state is unstable for $a<a_\text{c}$. Numerical continuation is employed on a one-dimensional domain of length $L=100\pi$ with Neumann boundary conditions. As a consequence only modes with wavenumbers $k_n= n \pi/L=n/100$ are selected.
\begin{figure}
\includegraphics[width=\hsize]{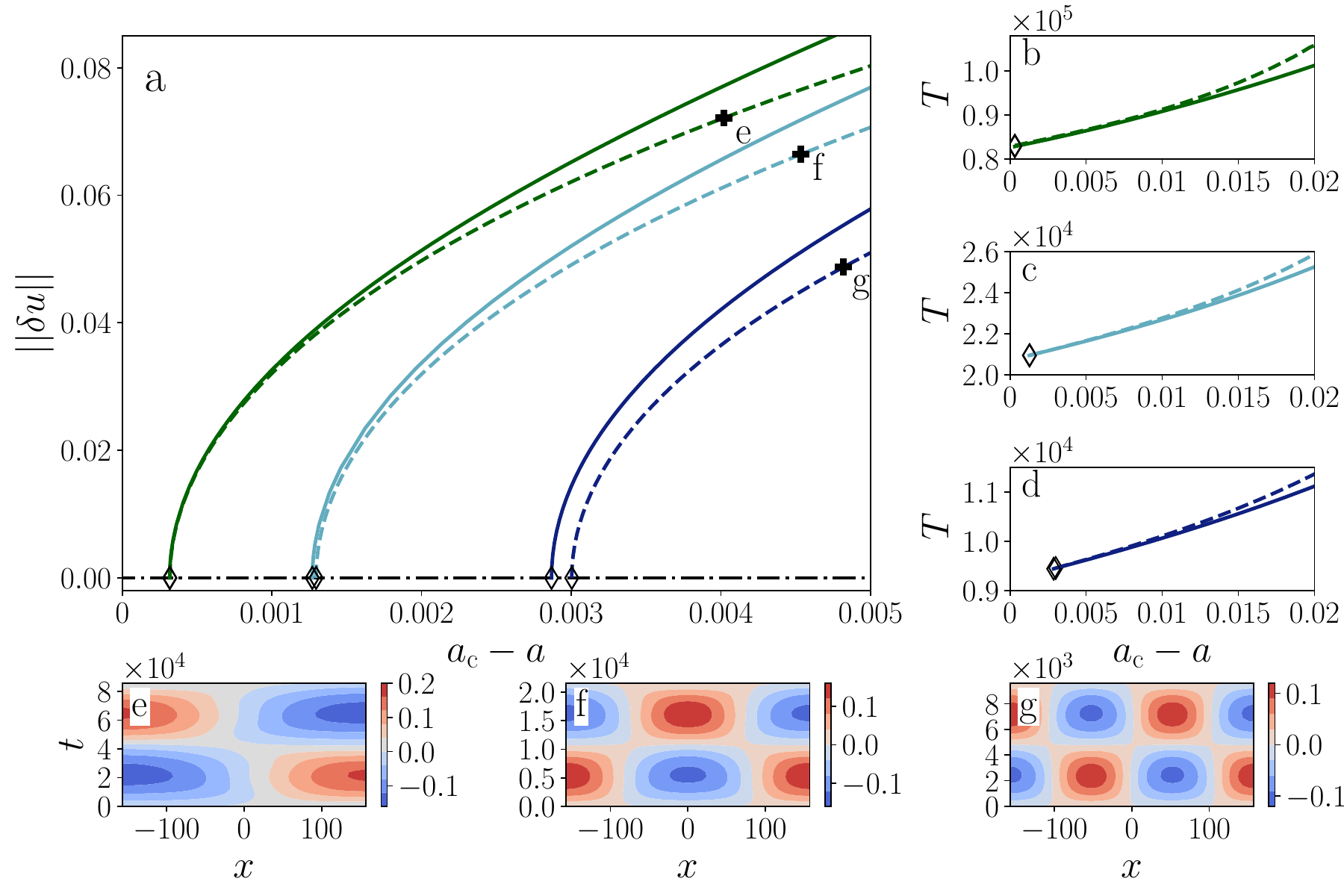}
\caption{Quantitative comparison of the linear and nonlinear
    behavior exhibited by the three-component reaction-diffusion
    system~\eqref{eq:RD3_rewritten} (solid lines in (a)-(d)) and the
    nonreciprocal Cahn-Hilliard equations~\eqref{eq:general_nonCH_A}
    with the derived coefficients given by Eq.~\eqref{eq:coeff}
    (dashed lines in (a)-(d)). The bifurcation diagram in panel (a)
    shows the time-averaged norm $||\delta u||$ \eqref{eq:norm} as a
    function of $a_\text{c}-a$. The horizontal dot-dashed line
    represents the trivial uniform state that becomes unstable in
    three primary Hopf bifurcations indicated by diamond
    symbols. There, branches of nonlinear standing waves of respective
    spatial wavenumber $k_1$, $k_2$ and $k_3$ emerge (starting from
    the left). Respective
    examples of $\rho(x,t)$ are shown as space-time plots in panels~(e)-(g) at
    parameters indicated by plus symbols in (a). Panels~(b)-(d) 
    compare the period $T$ of these states in a larger $a$-range.
}\label{fig:comparison}
\end{figure}
Fig.~\ref{fig:comparison}~(a) shows the resulting bifurcation diagram
in the vicinity of the first three primary Hopf bifurcations that are
indicated by diamond symbols, i.e., the parameter region where the
trivial state (black dot-dashed line) subsequently becomes unstable
with respect to the modes $k_1$, $k_2$ and $k_3$. As a solution
measure we employ the time-averaged norm
\begin{equation}\label{eq:norm}
||\delta u||= \frac{1}{T}\int_0^T {\rm d} t \sqrt{\frac{1}{L} \int_0^L  \left( \rho(x,t)^2 +\sigma(x,t)^2\right) {\rm d} x }\,,
\end{equation}
where $T$ is the period of the corresponding time-periodic state.  The emerging branches of standing waves are illustrated by dashed and solid lines for the amplitude and reaction-diffusion equations, respectively. It shows that the amplitude equation provides quite a good quantitative approximation of the reaction-diffusion system close to the onset of the conserved-Hopf instability. The approximation becomes exact at $a=a_\text{c}$ where the band of linear growing modes occur at zero wavenumber.  In consequence, for the employed finite system the
  primary bifurcations related to the mode of wavenumber $k_1$ in the
  full and approximated system are very close to each other (a small
  deviation in Fig.~\ref{fig:comparison}~(a) is not visible to the
  eye). The deviation is larger for the second and third primary bifurcations. Similarly, the deviations between the corresponding branches increase with the distance to $a=a_\text{c}$.  Panels~(e)-(g) respectively illustrate via space-time plots the qualitatively different standing waves on the three branches at the parameter values indicated by the bold plus symbols in panel~(a). Note that we only show the standing waves exhibited by the amplitude equation, the corresponding states of the reaction-diffusion system are very similar. The three states in (e)-(g) differ by their spatial periodicity and by their temporal period $T$. The former is a consequence of the large-scale property of the instability, hence, the nonlinear states inherit the spatial periodicity of the linear modes $k_1$ to $k_3$ that gain a positive growth rate at the respective primary bifurcation. The period is directly related to the behavior of the frequency, $\Omega(k) \sim k^2$, in the linear stability regime for $k\to 0$. That is, the time period close to the bifurcation behaves as $T\sim 1/k^2$, the qualitative behavior is inherited by the nonlinear states.  This is illustrated in panels~(b)-(d) where $T$ is plotted as a function of $a_\text{c}-a$ for the full and approximated model - a good agreement is found for $a_\text{c}-a\leq 0.02$.

Overall, Fig.~\ref{fig:comparison} shows that the amplitude equation provides a rather good quantitative estimate of the full dynamics.  In other words, the presented example verifies that the general nonreciprocal Cahn-Hilliard model can be employed as a faithful amplitude equation close to a conserved-Hopf instability.  In particular, for the supercritically emerging branches considered here, the amplitude equation correctly predicts the regular oscillations of low amplitude that would also be observed as the final oscillatory state in a time simulation of the original model. For subcritical cases one might need to go to higher orders in the weakly nonlinear expansion to faithfully predict the developing large-amplitude oscillations, although the local quadratic terms in the potentials in the derived Cahn-Hilliard model can, in principle, account for this. This is similar to other linear instabilities and the corresponding amplitude equations. However, the power of the description via amplitude equations lies in predicting qualitative changes in the weakly nonlinear behavior, such as the occurrence of secondary bifurcations, which are paradigmatic for models with a conserved-Hopf instability.  Further note that for our present example the two conservation laws reduce the local phase space (defined as in Ref.~\cite{BrHF2020prx}) to solely one dimension. Therefore, the emerging behavior will be less complex than expected for more complicated cases, e.g., the Min oscillations modeled in Ref.~\cite{HaFr2018np} by a reaction-diffusion system with two conservation laws or the liquid layer covered by active surfactants modeled in Ref.~\cite{PoTS2016epje} by a coupled thin-film and Smoluchowski equation also with two conservation laws.
More complex systems with conserved and nonconserved order parameter fields may exhibit both, conserved-Hopf and Hopf instabilities. Their nonlinear interaction may result in intricate chaotic behavior which may not be captured by the derived amplitude equation. Nevertheless, it is still likely that properties of secondary instabilities captured by the nonreciprocal Cahn-Hilliard equation will indicate how chaotic behavior can arise.  However, here, we have not considered the stability properties of the emerging standing wave states obtained with the amplitude equation and leave further studies of the qualitative behavior for the future.


\begin{thebibliography}{70}
\providecommand{\natexlab}[1]{#1}
\providecommand{\url}[1]{\texttt{#1}}
\expandafter\ifx\csname urlstyle\endcsname\relax
  \providecommand{\doi}[1]{doi: #1}\else
  \providecommand{\doi}{doi: \begingroup \urlstyle{rm}\Url}\fi

\bibitem[Fang et~al.(2019)Fang, Kruse, Lu, and Wang]{FKLW2019rmp}
X.~Fang, K.~Kruse, T.~Lu, and J.~Wang.
\newblock Nonequilibrium physics in biology.
\newblock \emph{Rev. Mod. Phys.}, 91:\penalty0 045004, 2019.
\newblock \doi{10.1103/revmodphys.91.045004}.

\bibitem[Marchetti et~al.(2013)Marchetti, Joanny, Ramaswamy, Liverpool, Prost,
  Rao, and Simha]{MJRL2013rmp}
M.~C. Marchetti, J.~F. Joanny, S.~Ramaswamy, T.~B. Liverpool, J.~Prost, M.~Rao,
  and R.~A. Simha.
\newblock Hydrodynamics of soft active matter.
\newblock \emph{Rev. Mod. Phys.}, 85:\penalty0 1143--1189, 2013.
\newblock \doi{10.1103/RevModPhys.85.1143}.

\bibitem[Wang et~al.(2021)Wang, Qian, and Xu]{WaQX2021sm}
H.~Wang, T.~Qian, and X.~Xu.
\newblock Onsager's variational principle in active soft matter.
\newblock \emph{Soft Matter}, 2021.
\newblock \doi{10.1039/d0sm02076a}.

\bibitem[Shaebani et~al.(2020)Shaebani, Wysocki, Winkler, Gompper, and
  Rieger]{SWWG2020nrp}
M.~R. Shaebani, A.~Wysocki, R.~G. Winkler, G.~Gompper, and H.~Rieger.
\newblock Computational models for active matter.
\newblock \emph{Nat. Rev. Phys.}, 2:\penalty0 181--199, 2020.
\newblock \doi{10.1038/s42254-020-0152-1}.

\bibitem[Bowick et~al.(2022)Bowick, Fakhri, Marchetti, and
  Ramaswamy]{BFMR2022prx}
M.~J. Bowick, N.~Fakhri, M.~C. Marchetti, and S.~Ramaswamy.
\newblock Symmetry, thermodynamics, and topology in active matter.
\newblock \emph{Phys. Rev. X}, 12:\penalty0 010501, 2022.
\newblock \doi{10.1103/PhysRevX.12.010501}.

\bibitem[Mikhailov(1999)]{Mikhailov1999}
A.~S. Mikhailov.
\newblock \emph{Foundations of Synergetics {I}: {D}istributed active systems}.
\newblock Springer Verlag, Berlin, 1999.

\bibitem[Purwins et~al.(2010)Purwins, B{\"o}deker, and
  Amiranashvili]{PuBA2010ap}
H.~G. Purwins, H.~U. B{\"o}deker, and S.~Amiranashvili.
\newblock Dissipative solitons.
\newblock \emph{Adv. Phys.}, 59:\penalty0 485--701, 2010.
\newblock \doi{10.1080/00018732.2010.498228}.

\bibitem[Liehr(2013)]{Liehr2013}
A.~Liehr.
\newblock \emph{Dissipative Solitons in Reaction Diffusion Systems: Mechanisms,
  Dynamics, Interaction}.
\newblock Springer Series in Synergetics. Springer Berlin Heidelberg, 2013.
\newblock ISBN 9783642312519.

\bibitem[Konow et~al.(2021)Konow, Dolnik, and Epstein]{KoDE2021ptrsapes}
C.~Konow, M.~Dolnik, and I.~R. Epstein.
\newblock Insights from chemical systems into {T}uring-type morphogenesis.
\newblock \emph{Philos. Trans. R. Soc. A-Math. Phys. Eng. Sci.}, 379:\penalty0
  20200269, 2021.
\newblock \doi{10.1098/rsta.2020.0269}.

\bibitem[Berry et~al.(2018)Berry, Brangwynne, and Haataja]{BeBH2018rpp}
J.~Berry, C.~P. Brangwynne, and M.~Haataja.
\newblock Physical principles of intracellular organization via active and
  passive phase transitions.
\newblock \emph{Rep. Prog. Phys.}, 80:\penalty0 046601, 2018.
\newblock \doi{10.1088/1361-6633/aaa61e}.

\bibitem[Chen and Kolokolnikov(2014)]{ChKo2014jrsi}
Y.~X. Chen and T.~Kolokolnikov.
\newblock A minimal model of predator-swarm interactions.
\newblock \emph{J. R. Soc. Interface}, 11:\penalty0 20131208, 2014.
\newblock \doi{10.1098/rsif.2013.1208}.

\bibitem[Ivlev et~al.(2015)Ivlev, Bartnick, Heinen, Du, Nosenko, and
  L{\"o}wen]{IBHD2015prx}
A.~V. Ivlev, J.~Bartnick, M.~Heinen, C.~R. Du, V.~Nosenko, and H.~L{\"o}wen.
\newblock Statistical mechanics where {N}ewton's third law is broken.
\newblock \emph{Phys. Rev. X}, 5:\penalty0 011035, 2015.
\newblock \doi{10.1103/PhysRevX.5.011035}.

\bibitem[Kryuchkov et~al.(2018)Kryuchkov, Ivlev, and Yurchenko]{KrIY2018sm}
N.~P. Kryuchkov, A.~V. Ivlev, and S.~O. Yurchenko.
\newblock Dissipative phase transitions in systems with nonreciprocal effective
  interactions.
\newblock \emph{Soft Matter}, 14:\penalty0 9720--9729, 2018.
\newblock \doi{10.1039/c8sm01836g}.

\bibitem[Nasouri and Golestanian(2020)]{NaGo2020prl}
B.~Nasouri and R.~Golestanian.
\newblock Exact phoretic interaction of two chemically active particles.
\newblock \emph{Phys. Rev. Lett.}, 124:\penalty0 168003, 2020.
\newblock \doi{10.1103/physrevlett.124.168003}.

\bibitem[Fruchart et~al.(2021)Fruchart, Hanai, Littlewood, and
  Vitelli]{FHLV2021n}
M.~Fruchart, R.~Hanai, P.~B. Littlewood, and V.~Vitelli.
\newblock Non-reciprocal phase transitions.
\newblock \emph{Nature}, 592:\penalty0 363--369, 2021.
\newblock \doi{10.1038/s41586-021-03375-9}.

\bibitem[You et~al.(2020)You, Baskaran, and Marchetti]{YoBM2020pnasusa}
Z.~H. You, A.~Baskaran, and M.~C. Marchetti.
\newblock Nonreciprocity as a generic route to traveling states.
\newblock \emph{Proc. Natl. Acad. Sci. U. S. A.}, 117:\penalty0 19767--19772,
  2020.
\newblock \doi{10.1073/pnas.2010318117}.

\bibitem[Saha et~al.(2020)Saha, Agudo-Canalejo, and Golestanian]{SaAG2020prx}
S.~Saha, J.~Agudo-Canalejo, and R.~Golestanian.
\newblock Scalar active mixtures: {T}he non-reciprocal {C}ahn-{H}illiard model.
\newblock \emph{Phys. Rev. X}, 10:\penalty0 041009, 2020.
\newblock \doi{10.1103/PhysRevX.10.041009}.

\bibitem[Frohoff-H\"ulsmann et~al.(2021)Frohoff-H\"ulsmann, Wrembel, and
  Thiele]{FrWT2021pre}
T.~Frohoff-H\"ulsmann, J.~Wrembel, and U.~Thiele.
\newblock Suppression of coarsening and emergence of oscillatory behavior in a
  {C}ahn-{H}illiard model with nonvariational coupling.
\newblock \emph{Phys. Rev. E}, 103:\penalty0 042602, 2021.
\newblock \doi{10.1103/PhysRevE.103.042602}.

\bibitem[Cahn(1965)]{Cahn1965jcp}
J.~W. Cahn.
\newblock Phase separation by spinodal decomposition in isotropic systems.
\newblock \emph{J. Chem. Phys.}, 42:\penalty0 93--99, 1965.
\newblock \doi{10.1063/1.1695731}.

\bibitem[Hohenberg and Halperin(1977)]{HoHa1977rmp}
P.~C. Hohenberg and B.~I. Halperin.
\newblock Theory of dynamic critical phenomena.
\newblock \emph{Rev. Mod. Phys.}, 49:\penalty0 435--479, 1977.
\newblock \doi{10.1103/RevModPhys.49.435}.

\bibitem[Nauman and He(2001)]{NaHe2001ces}
E.~B. Nauman and D.~Q. He.
\newblock Nonlinear diffusion and phase separation.
\newblock \emph{Chem. Eng. Sci.}, 56:\penalty0 1999--2018, 2001.
\newblock \doi{10.1016/S0009-2509(01)00005-7}.

\bibitem[Mao et~al.(2019)Mao, Kuldinow, Haataja, and Ko{\v{s}}mrlj]{MKHK2019sm}
S.~Mao, D.~Kuldinow, M.~P. Haataja, and A.~Ko{\v{s}}mrlj.
\newblock Phase behavior and morphology of multicomponent liquid mixtures.
\newblock \emph{Soft Matter}, 15:\penalty0 1297--1311, 2019.
\newblock \doi{10.1039/c8sm02045k}.

\bibitem[Frohoff-H\"ulsmann and Thiele(2021)]{FrTh2021ijam}
T.~Frohoff-H\"ulsmann and U.~Thiele.
\newblock Localised states in coupled {C}ahn-{H}illiard equations.
\newblock \emph{IMA J. Appl. Math.}, 86:\penalty0 924--943, 2021.
\newblock \doi{10.1093/imamat/hxab026}.

\bibitem[Hoyle(2006)]{Hoyle2006}
R.~B. Hoyle.
\newblock \emph{Pattern Formation -- An Introduction to Methods}.
\newblock Cambridge University Press, Cambridge, 2006.
\newblock \doi{10.1017/CBO9780511616051}.

\bibitem[Cross and Hohenberg(1993)]{CrHo1993rmp}
M.~C. Cross and P.~C. Hohenberg.
\newblock Pattern formation out of equilibrium.
\newblock \emph{Rev. Mod. Phys.}, 65:\penalty0 851--1112, 1993.
\newblock \doi{10.1103/RevModPhys.65.851}.

\bibitem[Aranson and Kramer(2002)]{ArKr2002rmp}
I.~S. Aranson and L.~Kramer.
\newblock The world of the complex {G}inzburg-{L}andau equation.
\newblock \emph{Rev. Mod. Phys.}, 74:\penalty0 99--143, 2002.
\newblock \doi{10.1103/revmodphys.74.99}.

\bibitem[Matthews and Cox(2000)]{MaCo2000n}
P.~C. Matthews and S.~M. Cox.
\newblock Pattern formation with a conservation law.
\newblock \emph{Nonlinearity}, 13:\penalty0 1293--1320, 2000.
\newblock \doi{10.1088/0951-7715/13/4/317}.

\bibitem[Winterbottom et~al.(2005)Winterbottom, Matthews, and Cox]{WiMC2005n}
D.~M. Winterbottom, P.~C. Matthews, and S.~M. Cox.
\newblock Oscillatory pattern formation with a conserved quantity.
\newblock \emph{Nonlinearity}, 18:\penalty0 1031--1056, 2005.
\newblock \doi{10.1088/0951-7715/18/3/006}.

\bibitem[Beta et~al.(2020)Beta, Gov, and Yochelis]{BeGY2020c}
C.~Beta, N.~S. Gov, and A.~Yochelis.
\newblock Why a large-scale mode can be essential for understanding
  intracellular actin waves.
\newblock \emph{Cells}, 9:\penalty0 1533, 2020.
\newblock \doi{10.3390/cells9061533}.

\bibitem[Yochelis et~al.(2022)Yochelis, Flemming, and Beta]{YoFB2022prl}
A.~Yochelis, S.~Flemming, and C.~Beta.
\newblock Versatile patterns in the actin cortex of motile cells:
  Self-organized pulses can coexist with macropinocytic ring-shaped waves.
\newblock \emph{Phys. Rev. Lett.}, 129:\penalty0 088101, 2022.
\newblock \doi{10.1103/physrevlett.129.088101}.

\bibitem[Thiele et~al.(2013)Thiele, Archer, Robbins, Gomez, and
  Knobloch]{TARG2013pre}
U.~Thiele, A.~J. Archer, M.~J. Robbins, H.~Gomez, and E.~Knobloch.
\newblock Localized states in the conserved {S}wift-{H}ohenberg equation with
  cubic nonlinearity.
\newblock \emph{Phys. Rev. E}, 87:\penalty0 042915, 2013.
\newblock \doi{10.1103/PhysRevE.87.042915}.

\bibitem[Knobloch(2016)]{Knob2016ijam}
E.~Knobloch.
\newblock Localized structures and front propagation in systems with a
  conservation law.
\newblock \emph{IMA J. Appl. Math.}, 81:\penalty0 457--487, 2016.
\newblock \doi{10.1093/imamat/hxw029}.

\bibitem[Bray(1994)]{Bray1994ap}
A.~J. Bray.
\newblock Theory of phase-ordering kinetics.
\newblock \emph{Adv. Phys.}, 43:\penalty0 357--459, 1994.
\newblock \doi{10.1080/00018739400101505}.

\bibitem[Bergmann et~al.(2018)Bergmann, Rapp, and Zimmermann]{BeRZ2018pre}
F.~Bergmann, L.~Rapp, and W.~Zimmermann.
\newblock Active phase separation: a universal approach.
\newblock \emph{Phys. Rev. E}, 98:\penalty0 020603, 2018.
\newblock \doi{10.1103/PhysRevE.98.020603}.

\bibitem[Ishihara et~al.(2007)Ishihara, Otsuji, and Mochizuki]{IsOM2007pre}
S.~Ishihara, M.~Otsuji, and A.~Mochizuki.
\newblock Transient and steady state of mass-conserved reaction-diffusion
  systems.
\newblock \emph{Phys. Rev. E}, 75:\penalty0 015203, 2007.
\newblock \doi{10.1103/PhysRevE.75.015203}.

\bibitem[Otsuji et~al.(2007)Otsuji, Ishihara, Co, Kaibuchi, Mochizuki, and
  Kuroda]{OICK2007pcb}
M.~Otsuji, S.~Ishihara, C.~Co, K.~Kaibuchi, A.~Mochizuki, and S.~Kuroda.
\newblock A mass conserved reaction-diffusion system captures properties of
  cell polarity.
\newblock \emph{PLoS Comput. Biol.}, 3:\penalty0 e108, 2007.
\newblock \doi{10.1371/journal.pcbi.0030108}.

\bibitem[Ei et~al.(2012)Ei, Izuhara, and Mimura]{EiIM2012dcdsb}
S.~I. Ei, H.~Izuhara, and M.~Mimura.
\newblock Infinite dimensional relaxation oscillation in aggregation-growth
  systems.
\newblock \emph{Discrete Contin. Dyn. Syst.-Ser. B}, 17:\penalty0 1859--1887,
  2012.
\newblock \doi{10.3934/dcdsb.2012.17.1859}.

\bibitem[Wettmann et~al.(2014)Wettmann, Bonny, and Kruse]{WeBK2014if}
L.~Wettmann, M.~Bonny, and K.~Kruse.
\newblock Effects of molecular noise on bistable protein distributions in
  rod-shaped bacteria.
\newblock \emph{Interface Focus}, 4:\penalty0 20140039, 2014.
\newblock \doi{10.1098/rsfs.2014.0039}.

\bibitem[Bernitt et~al.(2017)Bernitt, Dobereiner, Gov, and
  Yochelis]{BDGY2017nc}
E.~Bernitt, H.~G. Dobereiner, N.~S. Gov, and A.~Yochelis.
\newblock Fronts and waves of actin polymerization in a bistability-based
  mechanism of circular dorsal ruffles.
\newblock \emph{Nat. Commun.}, 8:\penalty0 15863, 2017.
\newblock \doi{10.1038/ncomms15863}.

\bibitem[Chiou et~al.(2018)Chiou, Ramirez, Elston, Witelski, Schaeffer, and
  Lew]{CREW2018pcb}
J.~G. Chiou, S.~A. Ramirez, T.~C. Elston, T.~P. Witelski, D.~G. Schaeffer, and
  D.~J. Lew.
\newblock Principles that govern competition or co-existence in {Rho-GTPase}
  driven polarization.
\newblock \emph{PLoS Comput. Biol.}, 14:\penalty0 e1006095, 2018.
\newblock \doi{10.1371/journal.pcbi.1006095}.

\bibitem[Halatek and Frey(2018)]{HaFr2018np}
J.~Halatek and E.~Frey.
\newblock Rethinking pattern formation in reaction-diffusion systems.
\newblock \emph{Nature Phys.}, 14:\penalty0 507--514, 2018.
\newblock \doi{10.1038/s41567-017-0040-5}.

\bibitem[Brauns et~al.(2020)Brauns, Halatek, and Frey]{BrHF2020prx}
F.~Brauns, J.~Halatek, and E.~Frey.
\newblock Phase-space geometry of mass-conserving reaction-diffusion dynamics.
\newblock \emph{Phys. Rev. X}, 10:\penalty0 041036, 2020.
\newblock \doi{10.1103/PhysRevX.10.041036}.

\bibitem[Brauns et~al.(2021)Brauns, Weyer, Halatek, Yoon, and
  Frey]{BWHY2021prl}
F.~Brauns, H.~Weyer, J.~Halatek, J.~Yoon, and E.~Frey.
\newblock Wavelength selection by interrupted coarsening in reaction-diffusion
  systems.
\newblock \emph{Phys. Rev. Lett.}, 126:\penalty0 104101, 2021.
\newblock \doi{10.1103/physrevlett.126.104101}.

\bibitem[Rapp and Zimmermann(2019)]{RaZi2019pre}
L.~Rapp and W.~Zimmermann.
\newblock Universal aspects of collective behavior in chemotactic systems.
\newblock \emph{Phys. Rev. E}, 100:\penalty0 032609, 2019.
\newblock \doi{10.1103/PhysRevE.100.032609}.

\bibitem[Bergmann and Zimmermann(2019)]{BeZi2019po}
F.~Bergmann and W.~Zimmermann.
\newblock On system-spanning demixing properties of cell polarization.
\newblock \emph{PLoS One}, 14:\penalty0 e0218328, 2019.
\newblock \doi{10.1371/journal.pone.0218328}.

\bibitem[Zimmermann(1997)]{Zimm1997pa}
W.~Zimmermann.
\newblock Stability of traveling waves for a conserved field.
\newblock \emph{Physica A}, 237:\penalty0 405--412, 1997.
\newblock \doi{10.1016/S0378-4371(96)00422-0}.

\bibitem[Wolansky(2002)]{Wola2002ejam}
G.~Wolansky.
\newblock Multi-components chemotactic system in the absence of conflicts.
\newblock \emph{Eur. J. Appl. Math.}, 13:\penalty0 641--661, 2002.
\newblock \doi{10.1017/S0956792501004843}.

\bibitem[Radszuweit et~al.(2013)Radszuweit, Alonso, Engel, and
  B{\"a}r]{RAEB2013prl}
M.~Radszuweit, S.~Alonso, H.~Engel, and M.~B{\"a}r.
\newblock Intracellular mechanochemical waves in an active poroelastic model.
\newblock \emph{Phys. Rev. Lett.}, 110:\penalty0 138102, 2013.
\newblock \doi{10.1103/PhysRevLett.110.138102}.

\bibitem[Pototsky et~al.(2016)Pototsky, Thiele, and Stark]{PoTS2016epje}
A.~Pototsky, U.~Thiele, and H.~Stark.
\newblock Mode instabilities and dynamic patterns in a colony of self-propelled
  surfactant particles covering a thin liquid layer.
\newblock \emph{Eur. Phys. J. E}, 39:\penalty0 1--19, 2016.
\newblock \doi{10.1140/epje/i2016-16051-4}.

\bibitem[John and B{\"a}r(2005)]{JoBa2005pb}
K.~John and M.~B{\"a}r.
\newblock Travelling lipid domains in a dynamic model for protein-induced
  pattern formation in biomembranes.
\newblock \emph{Phys. Biol.}, 2:\penalty0 123--132, 2005.
\newblock \doi{10.1088/1478-3975/2/2/005}.

\bibitem[Okada et~al.(2020)Okada, Sumino, Ito, and Kitahata]{OSIK2020pre}
M.~Okada, Y.~Sumino, H.~Ito, and H.~Kitahata.
\newblock Spontaneous deformation and fission of oil droplets on an aqueous
  surfactant solution.
\newblock \emph{Phys. Rev. E}, 102:\penalty0 042603, 2020.
\newblock \doi{10.1103/PhysRevE.102.042603}.

\bibitem[Waizumi et~al.(2021)Waizumi, Sakuta, Hayashi, Tsumoto, Takiguchi, and
  Yoshikawa]{WSH2021jcp}
T.~Waizumi, H.~Sakuta, M.~Hayashi, K.~Tsumoto, K.~Takiguchi, and K.~Yoshikawa.
\newblock Polymerization/depolymerization of actin cooperates with the
  morphology and stability of cell-sized droplets generated in a polymer
  solution under a depletion effect.
\newblock \emph{J. Chem. Phys.}, 155:\penalty0 075101, 2021.
\newblock \doi{10.1063/5.0055460}.

\bibitem[Chen et~al.(2017)Chen, Sadakane, Sakuta, Yao, and Yoshikawa]{CSS2017l}
Y.~J. Chen, K.~Sadakane, H.~Sakuta, C.~G. Yao, and K.~Yoshikawa.
\newblock Spontaneous oscillations and synchronization of active droplets on a
  water surface via {M}arangoni convection.
\newblock \emph{Langmuir}, 33:\penalty0 12362--12368, 2017.
\newblock \doi{10.1021/acs.langmuir.7b03061}.

\bibitem[Pototsky et~al.(2005)Pototsky, Bestehorn, Merkt, and
  Thiele]{PBMT2005jcp}
A.~Pototsky, M.~Bestehorn, D.~Merkt, and U.~Thiele.
\newblock Morphology changes in the evolution of liquid two-layer films.
\newblock \emph{J. Chem. Phys.}, 122:\penalty0 224711, 2005.
\newblock \doi{10.1063/1.1927512}.

\bibitem[Nepomnyashchy and Simanovskii(2017)]{NeSi2017pf}
A.~A. Nepomnyashchy and I.~B. Simanovskii.
\newblock Novel criteria for the development of monotonic and oscillatory
  instabilities in a two-layer film.
\newblock \emph{Phys. Fluids}, 29:\penalty0 092104, 2017.
\newblock \doi{10.1063/1.5001729}.

\bibitem[Nepomnyashchy and Shklyaev(2016)]{NeSh2016jpat}
A.~Nepomnyashchy and S.~Shklyaev.
\newblock Longwave oscillatory patterns in liquids: outside the world of the
  complex {G}inzburg-{L}andau equation.
\newblock \emph{J. Phys. A-Math. Theor.}, 49:\penalty0 053001, 2016.
\newblock \doi{10.1088/1751-8113/49/5/053001}.

\bibitem[Golovin et~al.(1997)Golovin, Nepomnyashchy, Pismen, and
  Riecke]{GNPR1997pd}
A.~A. Golovin, A.~A. Nepomnyashchy, L.~M. Pismen, and H.~Riecke.
\newblock Steady and oscillatory side-band instabilities in {M}arangoni
  convection with deformable interface.
\newblock \emph{Physica D}, 106:\penalty0 131--147, 1997.

\bibitem[Rottsch{\"a}fer and Doelman(1998)]{RoDo1998pd}
V.~Rottsch{\"a}fer and A.~Doelman.
\newblock On the transition from the {G}inzburg-{L}andau equation to the
  extended {F}isher-{K}olmogorov equation.
\newblock \emph{Physica D}, 118:\penalty0 261--292, 1998.
\newblock \doi{10.1016/S0167-2789(98)00035-9}.

\bibitem[De~Wit et~al.(1996)De~Wit, Lima, Dewel, and Borckmans]{DLDB1996pre}
A.~De~Wit, D.~Lima, G.~Dewel, and P.~Borckmans.
\newblock Spatiotemporal dynamics near a codimension-two point.
\newblock \emph{Phys. Rev. E}, 54:\penalty0 261--271, 1996.
\newblock \doi{10.1103/PhysRevE.54.261}.

\bibitem[Frohoff-H{\"u}lsmann et~al.(2023)Frohoff-H{\"u}lsmann, Holl, Knobloch,
  Gurevich, and Thiele]{FHKG2023pre}
T.~Frohoff-H{\"u}lsmann, M.~P. Holl, E.~Knobloch, S.~V. Gurevich, and
  U.~Thiele.
\newblock Stationary broken parity states in active matter models.
\newblock \emph{Phys. Rev. E}, 107:\penalty0 064210, 2023.
\newblock \doi{10.1103/PhysRevE.107.064210}.

\bibitem[Ma(2000)]{Ma2000jpsj}
Y.~Q. Ma.
\newblock Phase separation in ternary mixtures.
\newblock \emph{J. Phys. Soc. Jpn.}, 69:\penalty0 3597--3601, 2000.

\bibitem[Knobloch(1992)]{Knob1992}
E.~Knobloch.
\newblock Nonlocal amplitude equations.
\newblock In S.~Kai, editor, \emph{Pattern Formation in Complex Dissipative
  Systems}, pages 263--274, Singapur, 1992. World Scientific.
\newblock \doi{10.1142/9789814538039}.

\bibitem[Thiele et~al.(2016)Thiele, Archer, and Pismen]{ThAP2016prf}
U.~Thiele, A.~J. Archer, and L.~M. Pismen.
\newblock Gradient dynamics models for liquid films with soluble surfactant.
\newblock \emph{Phys. Rev. Fluids}, 1:\penalty0 083903, 2016.
\newblock \doi{10.1103/PhysRevFluids.1.083903}.

\bibitem[Jachalski et~al.(2014)Jachalski, Peschka, M{\"u}nch, and
  Wagner]{JPMW2014jem}
S.~Jachalski, D.~Peschka, A.~M{\"u}nch, and B.~Wagner.
\newblock Impact of interfacial slip on the stability of liquid two-layer
  polymer films.
\newblock \emph{J. Eng. Math.}, 86:\penalty0 9--29, 2014.
\newblock \doi{10.1007/s10665-013-9651-8}.

\bibitem[Dijkstra et~al.(2014)Dijkstra, Wubs, Cliffe, Doedel, Dragomirescu,
  Eckhardt, Gelfgat, Hazel, Lucarini, Salinger, Phipps, Sanchez-Umbria,
  Schuttelaars, Tuckerman, and Thiele]{DWCD2014ccp}
H.~A. Dijkstra, F.~W. Wubs, A.~K. Cliffe, E.~Doedel, I.~F. Dragomirescu,
  B.~Eckhardt, A.~Y. Gelfgat, A.~Hazel, V.~Lucarini, A.~G. Salinger, E.~T.
  Phipps, J.~Sanchez-Umbria, H.~Schuttelaars, L.~S. Tuckerman, and U.~Thiele.
\newblock Numerical bifurcation methods and their application to fluid
  dynamics: {A}nalysis beyond simulation.
\newblock \emph{Commun. Comput. Phys.}, 15:\penalty0 1--45, 2014.
\newblock \doi{10.4208/cicp.240912.180613a}.

\bibitem[Uecker et~al.(2014)Uecker, Wetzel, and Rademacher]{UeWR2014nmma}
H.~Uecker, D.~Wetzel, and J.~D.~M. Rademacher.
\newblock {pde2path} - a {Matlab} package for continuation and bifurcation in
  {2D} elliptic systems.
\newblock \emph{Numer. Math.-Theory Methods Appl.}, 7:\penalty0 58--106, 2014.
\newblock \doi{10.4208/nmtma.2014.1231nm}.

\bibitem[Pismen(2006)]{Pismen2006}
L.~M. Pismen.
\newblock \emph{Patterns and Interfaces in Dissipative Dynamics (Springer
  Series in Synergetics)}.
\newblock Springer-Verlag, Berlin Heidelberg, 2006.
\newblock ISBN 978-3-540-30431-9.
\newblock \doi{10.1007/3-540-30431-2}.

\bibitem[Oron and Nepomnyashchy(2004)]{OrNe2004pre}
A.~Oron and A.~A. Nepomnyashchy.
\newblock Long-wavelength thermocapillary instability with the {S}oret effect.
\newblock \emph{Phys. Rev. E}, 69:\penalty0 016313, 2004.
\newblock \doi{10.1103/PhysRevE.69.016313}.

\bibitem[Solon and Tailleur(2013)]{SoTa2013prl}
A.~P. Solon and J.~Tailleur.
\newblock Revisiting the flocking transition using active spins.
\newblock \emph{Phys. Rev. Lett.}, 111, 2013.
\newblock \doi{10.1103/physrevlett.111.078101}.

\bibitem[Solon and Tailleur(2015)]{SoTa2015pre}
A.~P. Solon and J.~Tailleur.
\newblock Flocking with discrete symmetry: the two-dimensional active {I}sing
  model.
\newblock \emph{Phys. Rev. E}, 92:\penalty0 042119, 2015.
\newblock \doi{10.1103/PhysRevE.92.042119}.

\end{thebibliography}
\end{document}